\documentclass[aps,pre,twocolumn,superscriptaddress,longbibliography ,floatfix]{revtex4-2}
\usepackage{graphicx,bm,amssymb,amsmath,dcolumn,hyperref}
\usepackage{multirow}
\usepackage{enumerate}
\usepackage{enumitem}
\usepackage{color}
\usepackage{subfigure}  
\usepackage{epstopdf}
\usepackage{bbm}
\usepackage{graphicx}
\usepackage{hyperref}
\usepackage{threeparttable}
\usepackage{amsthm}
\usepackage{mathtools}
\usepackage{color}
\usepackage{tikz}
\usepackage{float}
\usepackage{empheq}
\begin{document} 
\newcommand{\upcite}[1]{\textsuperscript{\textsuperscript{\cite{#1}}}}
\newcommand{\be}{\begin{equation}}
\newcommand{\ee}{\end{equation}}
\newcommand{\half}{\frac{1}{2}}
\newcommand{\ith}{^{(i)}}
\newcommand{\im}{^{(i-1)}}
\newcommand{\gae}
{\,\hbox{\lower0.5ex\hbox{$\sim$}\llap{\raise0.5ex\hbox{$>$}}}\,}
\newcommand{\lae}
{\,\hbox{\lower0.5ex\hbox{$\sim$}\llap{\raise0.5ex\hbox{$<$}}}\,}

\definecolor{blue}{rgb}{0,0,1}
\definecolor{red}{rgb}{1,0,0}
\definecolor{green}{rgb}{0,1,0}
\newcommand{\blue}[1]{\textcolor{blue}{#1}}
\newcommand{\red}[1]{\textcolor{red}{#1}}
\newcommand{\green}[1]{\textcolor{green}{#1}}
\newcommand{\orange}[1]{\textcolor{orange}{#1}}
\newcommand{\yd}[1]{\textcolor{blue}{#1}}

\newcommand{\scrA}{{\mathcal A}}
\newcommand{\scrB}{{\mathcal B}}
\newcommand{\scrE}{{\mathcal E}} 
\newcommand{\scrF}{{\mathcal F}} 
\newcommand{\scrL}{{\mathcal L}}
\newcommand{\scrH}{{\mathcal H}} 
\newcommand{\scrM}{{\mathcal M}} 
\newcommand{\scrN}{{\mathcal N}}
\newcommand{\scrS}{{\mathcal S}}
\newcommand{\scrs}{{\mathcal s}}
\newcommand{\scrP}{{\mathcal P}}
\newcommand{\scrO}{{\mathcal O}}
\newcommand{\scrQ}{{\mathcal Q}}
\newcommand{\scrR}{{\mathcal R}}
\newcommand{\scrC}{{\mathcal C}}
\newcommand{\scrV}{{\mathcal V}}
\newcommand{\scrD}{{\mathcal D}}
\newcommand{\scrG}{{\mathcal G}}
\newcommand{\scrZ}{{\mathcal Z}}
\newcommand{\scrW}{{\mathcal W}}
\newcommand{\scrU}{{\mathcal U}}
\newcommand{\Lm}{L_{\rm m}}
\newcommand{\ch}{\rm{chi}^2}
\newcommand{\diff}{{\rm d}}

\newcommand{\PP}{\mathbb{P}}
\newcommand{\ZZ}{\mathbb{Z}}
\newcommand{\EE}{\mathbb{E}}
\renewcommand{\d}{\mathrm{d}}
\newcommand{\dm}{d_{\rm min}}
\newcommand{\rhojunction}{\rho_{\rm j}}
\newcommand{\rhojunctionLim}{\rho_{{\rm j},0}}
\newcommand{\rhobranch}{\rho_{\rm b}}
\newcommand{\rhobranchLim}{\rho_{{\rm b},0}}
\newcommand{\rhononbridge}{\rho_{\rm n}}
\newcommand{\rhononbridgeLim}{\rho_{{\rm n},0}}
\newcommand{\percolationCluster}{C}
\newcommand{\leafFreeCluster}{C_{\rm \ell f}}
\newcommand{\bridgeFreeCluster}{C_{\rm bf}}
\newcommand{\df}{d_\textsc{f}}
\newcommand{\yt}{y_{\rm t}}
\newcommand{\yh}{y_{\rm h}}
\newcommand{\dB}{d_\textsc{B}}
\newcommand{\dfprime}{d'_{\rm f}}
\newcommand{\bfx}{{\bf x}}

\newcommand{\sC}{\mathcal{C}}
\newcommand{\LRa}[1]{\langle #1 \rangle}

\newcommand{\DLone}{D_\textsc{L1}}
\newcommand{\DLtwo}{D_\textsc{L2}}
\newcommand{\DFone}{D_\textsc{F1}}
\newcommand{\DFtwo}{D_\textsc{F2}}

\newcommand{\Dvf}{D_{\textsc{v1}}}
\newcommand{\Dff}{D_{\textsc{f1}}}
\newcommand{\Dfs}{D_\textsc{f2}}

\newcommand{\Dlf}{D_{\textsc{l1}}}
\newcommand{\Dls}{D_{\textsc{l2}}}
\newcommand{\dc}{d_{c}}

\title{Geometric scaling behaviors of the Fortuin-Kasteleyn Ising model in high dimensions }
\date{\today} 
\author{Sheng Fang}
\affiliation{MinJiang Collaborative Center for Theoretical Physics,
College of Physics and Electronic Information Engineering, Minjiang University, Fuzhou 350108, China}
\affiliation{Hefei National Research Center for Physical Sciences at the Microscales, University of Science and Technology of China, Hefei 230026, China}
	\author{Zongzheng Zhou}
	\email{eric.zhou@monash.edu}
	\affiliation{ARC Centre of Excellence for Mathematical and Statistical Frontiers (ACEMS),
		School of Mathematics, Monash University, Clayton, Victoria 3800, Australia}
\author{Youjin Deng}
\email{yjdeng@ustc.edu.cn}
\affiliation{MinJiang Collaborative Center for Theoretical Physics,
College of Physics and Electronic Information Engineering, Minjiang University, Fuzhou 350108, China}
\affiliation{Hefei National Research Center for Physical Sciences at the Microscales, University of Science and Technology of China, Hefei 230026, China}	
\affiliation{
Shanghai Research Center for Quantum Sciences, Shanghai 201315, China}

\begin{abstract}  
Recently, we argued [\href{https://iopscience.iop.org/article/10.1088/0256-307X/39/8/080502}{Chin. Phys. Lett. {\bf{39}}, 
080502 (2022)}] that the Ising model simultaneously exhibits two upper 
critical dimensions $(d_c=4, d_p=6)$ in the Fortuin-Kasteleyn (FK) random-cluster representation. 
In this paper, we perform a systematic study of the FK Ising model on hypercubic lattices 
with spatial dimensions $d$ from 5 to 7, and on the complete graph. 
We provide a detailed data analysis of the critical behaviors of a variety of quantities at and 
near the critical points. 
Our results clearly show that many quantities exhibit distinct critical phenomena for $4 < d < 6$ 
and $d\geq 6$, and thus strongly support the argument that $6$ is also an upper critical dimension. 
Moreover, for each studied dimension, 
we observe the existence of two configuration sectors, 
of two length scales as well as of two scaling windows, and thus,
two sets of critical exponents are needed to describe these behaviors. 
Our finding enriches the understanding of the critical phenomena in the Ising model.

\end{abstract}
\pacs{05.50.+q (lattice theory and statistics), 05.70.Jk (critical point phenomena),
64.60.F- (equilibrium properties near critical points, critical exponents)}
\maketitle

\section{Introduction}
\label{Introduction}
The Ising  model~\cite{friedli2017statistical} plays a fundamental role in statistical physics
and has an important influence on almost every branch of modern physics. 
The reduced Hamiltonian of the ferromagnetic Ising model without an external field is
        \begin{equation}
            \scrH = - K \sum_{\LRa{ij}} s_i s_j, \nonumber
        \end{equation}
where $s_i \in \{-1,1\}$ is the spin on vertex $i$, and the coupling strength $K>0$ 
acts as the inverse of temperature. 
The summation $\sum_{\LRa{ij}}$ is over all pairs of adjacent vertices. 
The Ising model was first proposed by Lenz in 1920 to explain the ferromagnetic phase 
transition~\cite{lenz1920beitrag}, and Ising showed that in one dimension (1D) 
there is no phase transition happening at any positive temperature~\cite{Ising1925Beitrag}. 
In 1944, the milestone was achieved by Onsager, who obtained the analytical expression of the free energy 
on the square lattice and discovered a continuous phase transition~\cite{Onsager1944Crystal}. 
The critical point is $K_c = \ln(1+\sqrt{2})/2$~\cite{Kramers1941statistics,Baxter2007} from duality arguments. 
The critical exponents $\beta$ and $\nu$, respectively characterizing the power-law behavior 
of the spontaneous magnetization and the divergence of correlation length near the critical point, 
are exactly known as $\beta = 1/8$~\cite{Yang1952Spontaneous} and $\nu = 1$. 
It has been proved that the Ising model exhibits a continuous phase transition on hypercubic lattices 
for all $d\geq 3$~\cite{AizenmanFernandez1986,AizenmanDuminilCopinVladas2015}. 
In 3D, only numerical estimates are available for both critical points and 
exponents~\cite{Deng2003Simultaneous,Ferrenberg2018Pushing,Hou2019Geometric,Guida19973dIsing}. 
In particular, the conformal bootstrap method~\cite{Kos2016Precision}, originating from high energy physics, 
has significantly improved the precision of critical exponents with the correlation-length 
exponent $\nu=0.629\,971(4)$ and susceptibility exponent $\gamma=1.237\,075(10)$. 
Renormalization group (RG) theory predicts that $d_c=4$ is the upper critical dimension for the Ising model, 
i.e., for $d\geq 4$ critical exponents take their mean-field values, 
e.g. $\beta = 1/2$ and $\nu=1/2$~\cite{FernandezFrohlichSokal13,AizenmanFernandez1986}.

Besides the spin representation, the Ising model has a well-known geometric representation, 
under the framework of the general $Q$-state random-cluster (RC) model~\cite{Grimmett2006Random} 
proposed by Fortuin and Kasteleyn (FK) in 1969. 
Given a graph $G \equiv (V,E)$ with vertex set $V$ and edge set $E$, 
the RC model is defined by choosing a spanning subgraph $(V,A)$ with a probability
\begin{equation}
    \pi(A) \propto v^{|A|}Q^{k(A)}\;. \nonumber
\end{equation}
Here $|A|$ is the number of edges on $A$, and $k(A)$ is the number of connected components (clusters) on $A$. 
Parameters $v$ and $Q$ are fugacity for edges and clusters. When $Q$ is an integer, 
the RC model can be mapped to the Potts model \cite{Wu1982The}, 
of which $(Q,v)=(2,e^{2K}-1)$ corresponds to the FK Ising model.

For $Q = 1$, the RC model reduces to the bond percolation model~\cite{Stauffer2018Introduction}. 
The RC model not only leads to many exact results in 2D, but it also provides a versatile platform to develop highly efficient cluster algorithms, such as the Wolff algorithm \cite{Wolff1989Collective}, the Swendsen-Wang (SW) algorithm \cite{Swendsen1987Nonuniversal}, the loop-cluster algorithm \cite{Zhang2020Loop}, etc.

It is natural to ask what the upper critical dimension of the RC model is. In the 1970s, it was suggested from RG analysis that for the general RC model, it could be either $4$ or $6$, depending on whether or not the $\phi^3$ term is taken into account in the field Hamiltonian~\cite{Zia1975Critical, Amit1976Renormalization}.

For percolation ($Q=1$) and the Ising model ($Q=2$), it is well known that their upper critical dimensions are $d_p = 6$~\cite{Chayes1987Upper} and $d_c = 4$, respectively. However, by studying the FK Ising model on the Bethe lattice and complete graph (CG), the authors in Ref.~\cite{Chayes1999Meanfield} conjectured that $6$ is the upper critical dimension of the FK Ising model, rather than 4. The CG is the graph on which all pairs of vertices are adjacent, and both the CG and Bethe lattice can be regarded as the $d \to \infty$ limit of lattices.

Recently, based on a combination of extensive simulations from $d = 4$ to 7, 
insights from RG theory, rigorous and numerical results on the CG, 
we argued that 
the FK Ising model has two upper critical dimensions $d_c=4$ and $d_p=6$, 
depending on which quantities to be considered~\cite{Fang2022Geometric}. 
We note that, to our knowledge, $d_p = 6$ cannot be seen from quantities in the spin representation. 
Compared with Ref.~\cite{Fang2022Geometric}, the goal of the current paper 
is to provide a systematic study of various quantities of the FK Ising model 
on high-dimensional (high-d) tori (lattices with periodic boundary conditions) 
and present a more detailed data analysis, such that the two-upper-critical-dimension phenomena 
is clearly demonstrated.
\par 

In the field of critical phenomena, the theory describing the asymptotic approach of finite systems to the thermodynamic limit near a continuous phase transition is called finite-size scaling (FSS). Thus, before moving on to our numerical results, let us briefly review some theoretical predictions to the FSS of the Ising model on high-d tori and boxes (lattices with free boundary conditions). The basic hypothesis of FSS is that the correlation length is cut off by the linear system size $L$, such that the singular part of the free energy density can be written as
\begin{equation}
    f(t,h) = L^{-d} \tilde{f}(tL^{y_t}, hL^{y_h})\;. \nonumber
\end{equation}
where $t = (K_c - K)/K_c$, $h$ is the magnetic scaling field, and exponents $y_t$ and $y_h$ are respectively the thermal and magnetic renormalization exponents. 
The scaling behaviors of many quantities can be derived accordingly. 
For example, the susceptibility $\chi$ corresponds to the second derivative of $f$ 
with respect to $h$, and scales as $\chi \sim L^{2 y_h - d}$ at the critical point $t=0$ 
and with zero field $h=0$. 
Above $d_c = 4$, the FSS of the Ising model is controlled by Gaussian fixed point (GFP) 
which gives $(y_t, y_h) = (2,1+d/2)$, that is, for $d\geq 4$ one expects $\chi \sim L^{2}$. However, it was observed that the FSS of $\chi$ above 4D depends on the boundary conditions~\cite{Brezin1985Finite,Binder1985Finitesize,Wittmann2014Finitesize,Flores-Sola2016Role,Grimm2017Geometric,Zhou2018Randomlength,Grimm2018Finitesize,Fang2020Complete,Lv2020Two,Camia2020effect}. The scaling $L^2$ was observed on boxes, but on tori, it was observed that $\chi \sim L^{d/2}$. \par 

To understand the behavior on tori, one can turn to the CG. The CG can be regarded as the mean-field approximation of models on high-d tori, since both of them are finite, translational invariant, and have large vertex degrees. 
However, models on the CG are often more tractable.
For the Ising model on the CG with volume $V$, it is known that $\chi \sim V^{1/2}\widetilde\chi(tV^{1/2})$ with $\widetilde\chi(\cdot)$ the scaling function. Let $V = L^d$ and use the scaling formula of $\chi$. Then one can see that the CG asymptotics predict two new exponents $(y^*_t, y^*_h) = (d/2, 3d/4)$ for high-d tori. Since spatial distance is not defined on the CG, it is reasonable to believe that spatial fluctuations are controlled by the GFP. In Ref.~\cite{Lv2020Two}, it is conjectured that one needs both the CG asymptotics and GFP to fully describe the FSS of the Ising model on high-d tori, and the free energy can be explicitly written as
\begin{equation}
\label{eq:free_energy}
f(t,h) = L^{-d}\tilde{f}_0(tL^{y_t},hL^{y_h}) + L^{-d} \tilde{f}_1(tL^{\yt^*}, hL^{\yh^*})\;.
\end{equation}
Predictions from Eq.~\eqref{eq:free_energy} for various quantities are all consistent with existing numerical results.

Although Eq.~\eqref{eq:free_energy} can describe very well the high-d-tori FSS of various quantities 
in the spin representation, the FSS in the geometric representation is still worth investigating,
since many geometric quantities have  no direct correspondence in the spin representation.
Before presenting results for the FK Ising model on the tori, let us first summarize the results 
known for the $d= \infty$ case, the CG. 
At the critical point and within the critical window of width ${\cal O}(V^{-1/2})$, 
it was proved that the size of the largest cluster $C_1 \sim V^{3/4}$; this again implies that $(y^*_t, y^*_h) = (d/2, 3d/4)$. For the second-largest cluster, it was proved 
that $C_2 = {\cal O}(\sqrt{V}\ln V)$~\cite{Luczak2006Phase}. Moreover, at $K = K_c - a V^{-1/3}$ with $a>0$, the system is in the percolation scaling window where $C_1, C_2 \sim V^{2/3}$, 
the same behavior as the two largest clusters in the critical percolation model on the CG~\cite{Bollobas1996randomcluster}. The Fisher exponent $\tau$, which characterizes the power-law decay of the cluster-number density, was conjectured to be $5/2$ in Ref.~\cite{Chayes1999Meanfield} and confirmed numerically in Ref.~\cite{Fang2021Percolation}. This is also consistent with the $\tau$ value for the critical percolation on the CG~\cite{Ben-Naim2005Kinetica}. Additionally, in Ref.~\cite{Fang2021Percolation}, we found that there is a vanishing sector in the whole configuration space, in which the scaling behaviors of all clusters follow the CG-percolation asymptotics, such as $C_1 \sim C_2 \sim V^{2/3}$ and $\tau = 5/2$. The probability of such a percolation sector vanishes with the rate of $V^{-1/12}$. The exponent $1/12$ happens to be the difference between the fractal dimensions of the largest cluster in the FK Ising model and the percolation model.

\begin{table}[b]
    \centering
    \begin{tabular}{|c|ll|ll|ll|}
    \hline
        $d$             & 5    &   & 6  & & 7 &   \\
    \hline
    $d_\textsc{l1}$   & $15/4 \,\,\,$ &$3.74(2)$   & $9/2 \,\,\,\,$ &$4.51(1)$   & $21/4\,\,\,$ & $5.18(2)$  \\
    $d_\textsc{f1}$   & $15/4$ &$3.76(1)$   & $9/2$ &$4.6(1)$    & $9/2$  & $4.55(12)$ \\
    $d_\textsc{l2}$   & $7/2$  &$3.486(11)$ & $4$   &$3.95(7)$   & $9/2$  & $4.48(3)$  \\
    $d_\textsc{f2}$   & $7/2$  &$3.61(3)$   & $4$   &$4.0(1)$    & $4$    & $4.1(1)$   \\
    \hline 
    \end{tabular}
    \caption{Conjectured values and numerical estimates for 
    the  finite-size ($d_\textsc{l1}$, $d_\textsc{l2}$) 
    and thermodynamic ($d_\textsc{f1}$, $d_\textsc{f2}$) fractal dimensions 
    for the largest and second-largest clusters. The theoretical values are $(d_\textsc{l1},d_\textsc{l2})=(3d/4, 1+d/2)$ for any $d \geq 4$, and $(d_\textsc{f1},d_\textsc{f2})=(d_\textsc{l1},d_\textsc{l2})$ for $6 > d \geq 4$ and $(9/2,4)$ for $d \geq 6$. }
    \label{tab:estimate_DFDL}
\end{table}

In this work, we present a systematic study to various quantities of the FK Ising model on tori with $d$ ranging from 5 to 7. We provide solid evidence to the existence of two length scales, two configuration sectors and two scaling windows, from which our conjecture that both $d_c = 4$ and $d_p = 6$ are the upper critical dimensions of the FK Ising model is supported. For $d=4$, since the theoretical prediction for the forms of logarithmic corrections of geometric quantities is incomplete, we leave the 4D case for future work. \par 
\paragraph*{Two length scales.}
We start with the fractal dimensions of clusters to demonstrate the existence of two length scales. 
Two finite-size fractal dimensions $d_\textsc{l1}, d_\textsc{l2}$ and two thermodynamic fractal dimensions $d_\textsc{f1}, d_\textsc{f2}$ are defined via $C_1 \sim L^{d_\textsc{l1}} \sim R_1^{d_\textsc{f1}}$ and $C_2 \sim L^{d_\textsc{l2}} \sim R_2^{d_\textsc{f2}}$. 
Here, $R_1$ and $R_2$ are the \emph{unwrapped} radii of gyration for $C_1$ and $C_2$, respectively, which represent the radii of the torus that clusters would have on the infinite lattice.
Our conjectured values for these fractal dimensions are summarized in Table~\ref{tab:estimate_DFDL}, and they are all consistent with numerical estimates. We find that $d_\textsc{l1}$ is consistent with $y^*_h = 3d/4$ for $d > 4$, following the CG asymptotics, and $d_\textsc{l2}$ is consistent with $1+d/2$, following the GFP predictions (possibly with multiplicative logarithmic corrections), which recovers the same leading behavior as the CG asymptotics ($V^{1/2}$) in the $d \to \infty$ limit. 
These results demonstrate that $4$ is an upper critical dimension, which is well-known in the spin representation, while for $d_\textsc{f1}$ and $d_\textsc{f2}$, we find that they are consistent with $d_\textsc{l1}$ and $d_\textsc{l2}$ for $d<6$. However, for $d \ge 6$, we find that $d_\textsc{f1}$ is consistent with $9/2$, and $d_\textsc{f2}$ is consistent with 4, independent of the spatial dimensions. We note that $4$ is the fractal dimension of percolation clusters on high-d lattices~\cite{Aharony1984Scaling}. These results suggest that $6$ is another upper critical dimension for the FK Ising model.

From these fractal dimensions, one can easily obtain the scaling behavior of the radius of gyration. 
For $d < 6$, both $R_1$ and $R_2$ are of order ${\cal O}(L)$. For $d \ge 6$, we have $R_1 \sim L^{d_\textsc{l1}/d_\textsc{f1}}=L^{d/6}$, consistent with that of the percolation model~\cite{Lu2023Finite}, and $R_2 \sim L^{d_\textsc{l2}/d_\textsc{f2}}=L^{1/4+d/8}$, both of which are larger than $L$. Therefore, the topology of these clusters changes at $d_p = 6$, namely, large clusters hardly wind around the torus when $d < 6$ but wind extensively when $d > 6$.

We next move to discuss the behavior of other clusters.
Our data show that for other clusters, their sizes $s$ scale with the radii $R$ as $s \sim R^{y_h}$ with $y_h = 1 + d/2$ for $4 < d < 6$ but $s\sim R^{4}$ for $d \geq 6$;
the latter is the behavior of percolation clusters on lattices with $d\geq 6$. 
The other interesting quantity to study is the cluster-number density $n(s)$ of these clusters, defined based on the fact that the number of clusters with size in $[s,s+\diff s)$ is $L^d n(s) \diff s$. 
It is typically expected that $n(s) \sim s^{-\tau}$ with a cutoff at $s$ close to the size of the largest cluster, and the hyperscaling relation $\tau=1+d/d_\textsc{l1}$ is believed to hold. This has been generally observed for percolation models in various dimensions~\cite{Huang2018Critical} and the FK Ising model below the upper critical dimension~\cite{Hou2019Geometric}. For $d > d_c$, since $C_1$ is much larger than $C_2$ and other clusters, it is plausible that the above scaling relation for $\tau$ fails. Indeed, for $4< d <6$, our numerics suggest that the Fisher exponent $\tau=1+\frac{d}{y_h}$ follows the GFP prediction. For $d \ge 6$, our data show that $\tau = 5/2$, which is consistent with the percolation model in high dimensions. Thus, the properties of other clusters in the FK Ising model follow the GFP prediction for $4< d < 6$, 
but exhibit percolation-cluster behavior for $d\geq 6$. 

From the fractal dimensions of other clusters and the cluster-number density, one can obtain the scaling of the number of spanning clusters $N_s$. A cluster is called spanning if its unwrapped extension  (defined in Sec.~\ref{simulation and observables}) exceeds the system size $L$. 
From the above discussions, one can see that the two largest clusters are spanning when $d > 6$. For other clusters, we have $s\sim R^4$ above 6D, and thus a cluster is spanning if its size is larger than ${\cal O}(L^4)$. It then follows that $N_s \sim L^d\int_{L^4} n(s)\diff s \sim$ $L^{d-6}$. Thus, $N_s$ is divergent when $d>6$ and possibly diverges logarithmically at $d=6$. We note that the scaling behaviors of $N_s$ for $d\geq 6 $ are the same as in the percolation model~\cite{Aizenman1997number,Fortunato2004Number}. By similar argument, one can obtain 
that $N_s \sim {\cal O}(1)$ for $d < 6$. The above scaling for $N_s$ is confirmed by our numerical data. \par
\paragraph*{Two configuration sectors.}
We then present evidence for the existence of two configuration sectors based on the size distribution of the largest cluster in the critical FK Ising model. 
Our data indicate that the distribution from finite-size systems converges to the limiting case quite slowly, i.e., a strong finite-size effect. Further investigation reveals that this is due to the existence of a special sector in the whole configuration space. Here, the sector is a set of bond configurations satisfying certain conditions, see Sec.~\ref{vanishing sector} for precise definitions. Interestingly, when conditioned on being in this sector, various quantities are observed to follow the GFP prediction for $4 < d < 6$; for example $C_1, C_2 \sim L^{1+d/2}$. For $d\geq 6$, quantities in this sector follow the high-d percolation behavior, like $C_1, C_2 \sim L^{2d/3}$. 
For all $d > 4$, the weight of the sector vanishes in the limit $L\rightarrow \infty$. Numerically, we observed that the vanishing rate is consistent with $L^{1-d/4}$ for $d<d_p$ and $L^{-d/12}$ for $d\ge d_p$; the latter is the same as on the CG \cite{Fang2021Percolation}. We note that for all $d>4$, the vanishing rate is equal to the difference between the finite-size fractal dimension of the largest cluster in the sector and in the whole configuration space.  \par 

\paragraph*{Two scaling windows.}
Finally, we present the existence of the two scaling windows  near the critical point.
For $d > d_c$, our data show that there is a critical window with a width of order ${\cal O}(L^{-d/2})$, consistent with the CG prediction that $y^*_t = d/2$. This is the FK Ising model at $K$ where $|K - K_c| = {\cal O}(L^{-d/2})$ exhibits the same scaling behavior as at the critical point $K_c$. Moreover, in the high-T regime, our data indicate the existence of another scaling window. For $d=5$, when $(K - K_c)L^{2}$ is constant, various quantities follow the GFP prediction, such that $C_1, C_2 \sim L^{1+d/2}$ and the radii $R_1, R_2$ are of constant order. For $d \ge 6$, we find that there is a percolation scaling window with a width of order ${\cal O}(L^{-d/3})$, i.e., when $(K - K_c)L^{d/3}$ is a constant, the FK clusters behave like percolation clusters. For example, one can observe $C_1, C_2 \sim L^{2d/3}$ and $R_1, R_2 \sim L^{d/6}$.

We then study the thermodynamic behavior of the two radii $R_1$ and $R_2$, which involve two correlation-length exponents $\nu_1$ and $\nu_2$ via $R_1 \sim |t|^{-\nu_1}$ as $t \to 0^{+}$ and $R_2 \sim |t|^{-\nu_2}$ as $t \to 0$. Based on the assumption of the standard FSS, one can recover the thermodynamic behavior from the FSS near $K_c$. As the critical point is approached from the high-T side ($t \rightarrow 0^+$), we find that $\nu_1 = \nu_2 = 1/2$ for all $d > 4$, consistent with the mean-field value for the correlation-length exponent. However, as $t\rightarrow 0^-$, we conjecture that $\nu_2=2/d$ for $4 < d < 6$ but $\nu_2 = 1/4 + 1/(2d)$ for $d \geq 6$. We note that in the $d \to \infty$ limit, one obtains $\nu_2 = 1/4$, which is consistent with the observation on the Bethe lattice~\cite{Chayes1999Meanfield}.


 
Finally, we note that these abundant phenomena of the FK Ising model cannot be observed within the spin representation. One possible reason is that many quantities in the geometric representation have no direct analog in the spin representation. Under the geometric representation, many spin quantities are decomposed into more refined geometric quantities, which exhibit deeper and more complex properties. For example, the susceptibility in the spin representation becomes the second moments of sizes of all clusters in the geometric representation, and obviously the latter contains much richer information. Indeed, by studying the behavior of these clusters, we found out that $d_p = 6$ is another upper critical dimension, i.e., clusters show many distinct behaviors below and above 6 dimensions.


The remainder of this article is organized as follows.  In Sec.~\ref{simulation and observables}, we provide the details of simulations and sampled quantities. Our numerical results are presented in Sec.~\ref{results1} and Sec.~\ref{results2}. Finally, we conclude with a discussion in Sec.~\ref{discussion}. 

\begin{table}[b]
    \centering
    \begin{tabular}{|l|p{2.5cm}|p{1.5cm} |p{1.5cm}|}
    \hline 
    $d$ &   $K_c$ &$V_{\rm max}$  &$N_{\rm sam}$  \\
    \hline 
         5& 0.113\,915\,0(4)  \cite{Blote1997Universality} & $51^5$   & $5 \times 10^5$ \\
         6& 0.092\,298\,2(3) \cite{Lundow2015Discontinuity} & $32^6$   & $2 \times 10^5$ \\
         7& 0.077\,708\,6(8) \cite{Lundow2015Discontinuity} & $20^7$   & $7 \times 10^4$ \\
         CG& $1/V$         & $2^{22}$ & $5 \times 10^6$ \\
    \hline 
    \end{tabular}
    \caption{The critical points $K_c$ and the largest simulated system volume $V_{\rm max}$ for $d=5,6,7$ and the CG. For each system, no less than $N_{\rm sam}$ independent samples are generated.}
    \label{tab:simulate_system}
\end{table}

\section{Simulation and Observable}
\label{simulation and observables}
We simulate the FK Ising model using a combination of the SW algorithm~\cite{Swendsen1987Nonuniversal} and the Wolff algorithm~\cite{Wolff1989Collective}. 
We use the SW algorithm to generate the FK cluster configuration, and between the consecutive SW steps, 
we use the Wolff algorithm to update the spin configurations, since it is believed that the Wolff algorithm  has a smaller dynamic exponent than the SW algorithm. We simulate the FK Ising model on high-d tori with $d=5,6,7$ and the CG. The critical points $K_c$, the largest system volume $V_{\rm 
max}$, and the number of independent samples $N_{\rm sam}$ are summarized 
in Table~\ref{tab:simulate_system}. 
In simulations, we sampled the following observables.
    \begin{enumerate}
        \item The size of the largest cluster $\scrC_1$ and the second-largest cluster $\scrC_2$;
        \item The number of clusters $\scrN(s)$ with size in the range $[s, s+\Delta s)$;
        \item For a cluster $C$, its radius of gyration $\scrR(C)$ is defined as 
            $$
                \scrR(C) = \sqrt{\sum_{u\in C} \frac{({\bf{x}}_u -\bar{{\bf x}})^2}{|C|} },
            $$
        where $\bar{\bf x} = \sum_{u \in C} {\bf{x}}_u/|C|$. Here ${\bf x}_u \in \ZZ^d$ is defined algorithmically as follows. First, choose the vertex, say, $o$, in $C$ with the smallest vertex label according to some fixed but arbitrary labeling. Set $x_o=0$. Start from the vertex $o$, and search through the cluster $C$ using breadth-first growth. Iteratively we set ${\bf x}_v = {\bf x}_u + {\bf e}_i$ if the vertex $v$ is traversed from $u$ along the $i$th direction, and set ${\bf x}_v = {\bf x}_u - {\bf e}_i$ if it is against the $i$th direction. Here ${\bf e}_i$ is the unit vector in the $i$th direction. The radii of the largest and second-largest clusters are denoted as $\scrR_1$ and $\scrR_2$, respectively;  
        \item The average radius of gyration of clusters with size in $[s, s+\Delta s)$ 
        $$
            \scrR(s) = \frac{ \sum_{C:|C| \in [s+\Delta s)} \scrR(C) }{  \scrN(s) }. 
        $$
        \item For each cluster $C$, we measure its unwrapped extension $\scrU$, which is the largest unwrapped distance in the first coordinate direction, i.e., $\scrU =  \text{max}_{u,v \in C} ({\bf x}_u - {\bf x}_v)_1$;
        \item The number of spanning clusters $\scrN_s$. A cluster is spanning if its $\scrU \geq L$.
    \end{enumerate}
    We choose $\Delta s$ properly to guarantee there are enough data for statistics in each interval.
    By taking the ensemble average $\LRa{\cdot}$ of these observables, we calculate the following quantities:
\begin{enumerate}
            \item The mean size of the largest cluster $C_1 = \LRa{\scrC_1}$ and the second-largest cluster $C_2 = \LRa{\scrC_2}$; 
            \item The radius of gyration $R(s) = \LRa{\scrR(s)}$ with a given cluster size $s$;
            \item The radius of gyration of the largest and second-largest clusters $R_1 = \LRa{\scrR_1}$ and $R_2 = \LRa{\scrR_2}$;
            \item The cluster-number density $n(s,V) = \frac{1}{V\Delta s} \LRa{\scrN(s)}$; 
            \item The number of spanning clusters $N_s = \LRa{\scrN_s}$. 
\end{enumerate}

\section{Results at criticality}
\label{results1}
We perform least-square fits on the FSS data. As a precaution against correction-to-scaling terms that we missed including in the fitting ansatz, we impose a lower cutoff $L \ge \Lm$ on the data points admitted in the fit, and we systematically study the effect on the residuals $\chi^2$ value by increasing $\Lm$. In general, the preferred fit for any given ansatz corresponds to the smallest $\Lm$ for which the goodness of  fit is reasonable and for which subsequent increases in $\Lm$ do not cause the $\chi^2$ value to drop by vastly more than one unit per degree of freedom. In practice, by “reasonable” we mean that $\chi^2/\rm{DF} \approx 1$, where DF is the number of degrees of freedom. The systematic error is estimated by comparing estimates from various sensible fitting ansatz.  \par  

For quantities without logarithmic corrections, we perform the least-square fits via the ansatz 
    \begin{equation}
    \label{eq:fitting_ansatz1}
        \scrO = L^{y_{\scrO}} (a_0 + b_1L^{y_1} + b_2 L^{y_2}) +c_0. 
    \end{equation}
For quantities with logarithmic corrections, we perform the least-square fits via the ansatz 
        \begin{equation}
        \label{eq:fitting_ansatz2}
            \scrO = L^{y_{\scrO}}  (\ln L +d_0)^{\hat{y}_{\scrO}}(a_0 + b_1L^{y_1} + b_2 L^{y_2} ) + c_0.
        \end{equation}
Here, we note $a_0L^{y_{\scrO}}$ describes the leading behavior of the quantities, $(\ln L +d_0)^{\hat{y}_{\scrO}}$ describes the logarithmic corrections, $b_1L^{y_1}$ and $b_2L^{y_2}$ describe the finite-size corrections with exponents $y_1, y_2$ less than 0, and $c_0$ originates from the 
 background contributions of various systems.

\begin{figure}[t]
\centering
\includegraphics[width=0.5\textwidth]{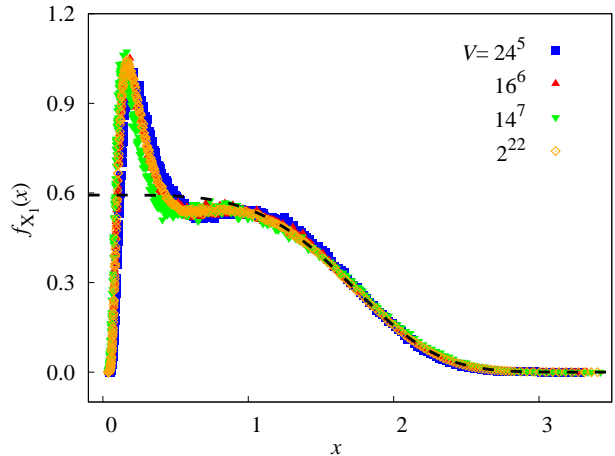}
\caption{Probability density functions of the rescaled size of the largest cluster $X_1 := \scrC_1/aV^{3/4}$. The factor $a$ is chosen to be $1.34, 1.25, 1.06, 1$ for $d=5,6,7$ and the CG, respectively. The dashed curve in the plot corresponds to Eq.~\eqref{eq:nsC1CG}, which is the $V\rightarrow \infty$ limiting case. As shown, results on finite-size systems are consistent with the limiting case only when $x \gtrsim 1$. The inconsistency part for small $x$ is due to the existence of an asymptotically vanishing sector in the configuration space, in which the scaling of $\scrC_1$ is not $V^{3/4}$.}         
\label{fig:nsC1t}
\end{figure}

\subsection{Existence of two sectors}
\label{two sectors}
\subsubsection{The probability distribution of the largest cluster}
\label{probability of C1}
In this section, we consider the probability distribution  of the largest cluster $f_{\scrC_1}(s)$.  Define $X_1 = \frac{\scrC_1}{aV^{3/4}}$ with a constant $a$ and its probability density function as $f_{X_1}(x)$. Then it follows that 
    \begin{equation}
       f_{\scrC_1}(s) ds = f_{X_1}(x) dx,
    \end{equation}
    where $dx = a^{-1}V^{-3/4} ds$ and $f_{X_1}(x)=a V^{3/4}f_{\scrC_1}(s)$. \par 

On the CG, it was proved in Ref.~\cite{Luczak2006Phase} that the probability density of $X_1$ with $a=1$ in the $V \to \infty$ limit follows 
    \begin{equation}
    \label{eq:nsC1CG}
        f_{X_1}^{\infty}(x) = \frac{ \exp(-x^4/12) }{ \int_0^{\infty} \exp(-t^4/12)dt}. 
    \end{equation}
Later, the authors in Ref.~\cite{Fang2021Percolation} confirmed it numerically. They further found that in finite volume $V$, the whole configuration space contains a percolation sector in which all clusters exhibit the same scaling behavior as the critical percolation on the CG. The probability of the percolation sector vanishes at a rate of order $V^{-1/12}$.  \par 

For $d > d_c=4$, it is believed that the scaling behavior of the FK Ising model obeys the CG asymptotics. 
In the study  on 5D tori, Lundow et al.~\cite{Lundow2015Complete} found that the probability distribution of FK clusters follows the CG asymptotics. 
In Fig.~\ref{fig:nsC1t}, we plot the probability distribution of the largest cluster $f_{X_1}(x)$ on high-d tori and CG. Similar to CG, it also has a double-peak distribution, and the first peak seems to disappear as system volume increases. We adjust the constant $a$ for various systems so that they have a good data collapse for $x \gtrsim 1$. The dashed line shows the CG prediction Eq.~\eqref{eq:nsC1CG}. This provides strong evidence that 4 is an upper critical dimension, since for $d>4$ it exhibits the same asymptotic scaling behavior as in the $V \to \infty$ limit. \par 

\begin{figure}[t]
    \centering
    \includegraphics[width=0.52\textwidth]{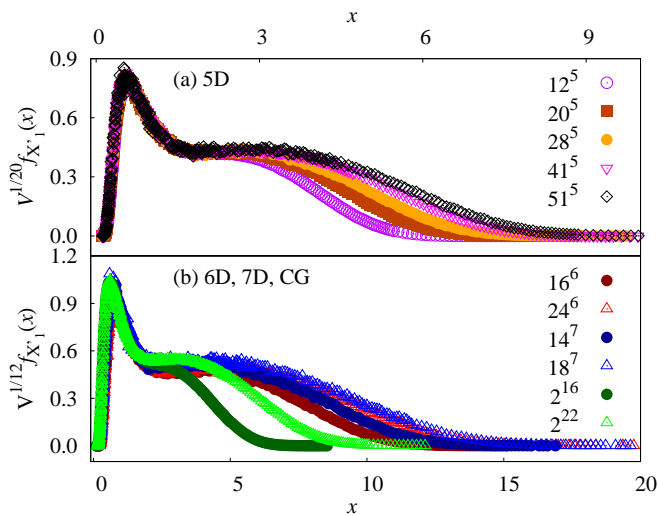}
    \caption{Demonstration of the vanishing sector, by plotting the probability density functions of the rescaled $\scrC_1$ for (a) $d=5$ with $X'_1=\scrC_1/L^{1+d/2}$ and (b) $d=6,7$, and CG with $X'_1=\scrC_1/V^{2/3}$. This strongly suggests that the sectors vanish with rate $V^{-1/20}$ for 5D and $V^{-1/12}$ for $d\geq 6$, and in the sectors the scaling of $C_1$ is respectively $L^{1+d/2}$ and $V^{2/3}$.}
    \label{fig:nsC1v}
\end{figure}

We then consider the scaling behavior of the probability density function when $\scrC_1$ is small. On the CG, a vanishing percolation sector was numerically observed  when $\scrC_1$ is small, and all clusters follow the CG-percolation asymptotics in this sector~\cite{Fang2021Percolation}.  
For high-d tori, one possible conjecture is that the scaling behavior of the vanishing sector should be consistent with CG for $d>4$. However, it may bring  a problem that $d=5$ is not sufficient to present the mean-field scaling behavior for percolation since its upper critical dimension is $6$. In Ref.~\cite{Fang2020Complete}, it was numerically observed that except for the largest clusters, all other clusters follow the GFP asymptotics. Therefore, we assume that the vanishing sector may follow the GFP asymptotics. 
Then, we define $X'_1=\scrC_1/L^{1+d/2}$ for $d=5$. Figure~\ref{fig:nsC1v}(a) plots $V^{1/20}f_{X'_1}(x)$ versus $x$, and it is clearly observed that it has a good data collapse for $x \lesssim 3$. We note that the term $V^{1/20} = L^{d/4-1}$ in 5D is from the quotient of the CG-Ising asymptotics $L^{3d/4}$ and the GFP asymptotics $L^{1+d/2}$. 
\par 

For $d \ge 6$, following the same procedure as $d=5$, we find that the data cannot be well collapsed. We then define $X'_1=\scrC_1/V^{2/3}$ and Fig.~\ref{fig:nsC1v}(b) shows the plot of $V^{1/12}f_{X'_1}(x)$ against $x$ on high-d tori with $d=6,7$ and CG. It can be numerically observed that when $x$ is small, the data collapse well for various systems. Here, we note that the exponent $1/12$ simply originates from the difference between the CG-Ising exponent $3/4$ and the CG-percolation exponent $2/3$. The good data collapse in Fig.~\ref{fig:nsC1v} implies  that there is a vanishing sector. The scaling of $\scrC_1$ in the sector, i.e. $L^{1+d/2}$ for $d=5$ and $V^{2/3}$ for $d \ge 6$, gives a hint that $6$ is also an upper critical dimension.   

\begin{figure}[t]
    \centering
    \includegraphics[width=0.5\textwidth]{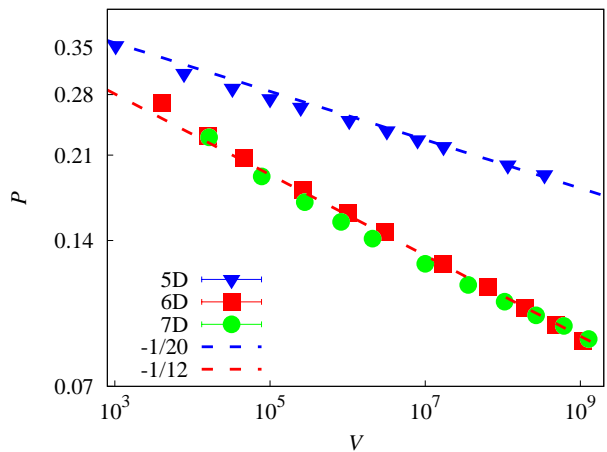}
    \caption{Plot to show that the vanishing rate of the vanishing sector is $V^{-1/20}$ for 5D and $V^{-1/12}$ for $d=6,7$. For $d\ge6$, the rate is consistent with the observation on the CG.}
    \label{fig:Pva}
\end{figure}

\subsubsection{The vanishing sector}
\label{vanishing sector}
We then consider the vanishing rate  and scaling behavior of clusters in the vanishing sector. 
The good data collapse in Fig.~\ref{fig:nsC1v} implies that there is an exponent $\theta$ and  positive constants $a_0$, $c_0$ so that 
        \begin{equation}
            \lim_{L \to \infty} V^{\theta} \PP\left[\frac{\scrC_1}{V_{\rm van}} \le a_0\right] = c_0
        \end{equation}
with $c_0 =\int_0^{a_0} f_{X_1'}(x)dx$. To precisely estimate the exponent $\theta$,
we set $a_0=1$ and choose  $V_{\rm van} = L^{1+d/2}$ for $d=5$ and $V_{\rm van}= V^{2/3}$ for $d \ge 6$, and count the probability $P$ of configurations with the size of the largest cluster $\scrC_1 \in (0,V_{\rm van}]$. 
We then perform the least-squares fits via the ansatz Eq.~\eqref{eq:fitting_ansatz1} with $L$ substituted by $V$. 
The final estimates of $\theta$ are 0.046(4) for $d=5$ and 0.088(3), 0.087(9) for $d=6,7$ respectively; the former for $d=5$ is consistent with the expected value $1/20$, and the latter for $d=6,7$ is consistent with $1/12$ within two standard deviations. 
Figure~\ref{fig:Pva} presents the probability $P$ versus system volume $V$ in the log-log scale for $d=5,6,7$ and the slopes of the data are well consistent with their expected value. \par

\begin{figure}[t]
    \centering
    \includegraphics[width=0.52\textwidth]{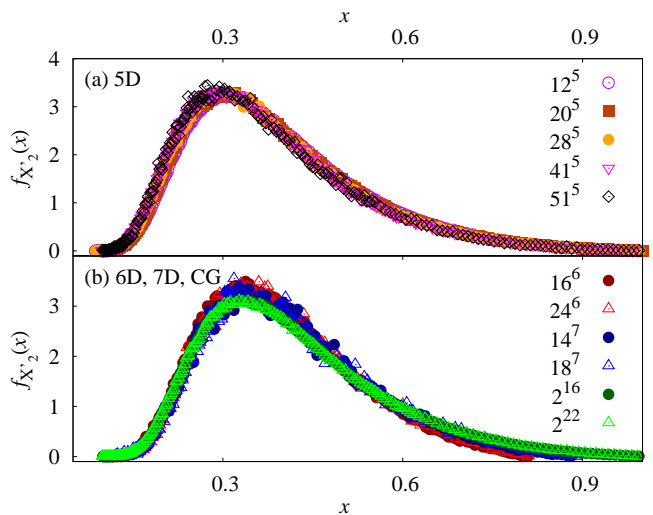}
    \caption{The probability density function of the rescaled second-largest cluster in the vanishing sector, which displays different scaling behaviors  for  (a) $d=5$, and (b) $d=6,7$ and CG. The variable $X'_2 := \scrC_2/L^{1+d/2}$ for $d=5$ and $X'_2 :=\scrC_2/(aV^{2/3})$ with $a=1.23, 1.12, 1$ for $d=6,7$ and CG, respectively.}
    \label{fig:nsC2v}
\end{figure}
We define the vanishing sector as the set of configurations with $ \scrC_1 \le V_{\rm van}$.
In this sector, it can be implied that the largest cluster $C_1 \sim L^{1+d/2}$ for $d=5$ and $C_1 \sim L^{2d/3}$ for $d=6,7$. We then consider the probability density of the second-largest cluster $\scrC_2$. 
We define the variable $X'_2 = \scrC_2/L^{1+d/2}$ for $d=5$ and $X'_2 = \scrC_2/(aV^{2/3})$ for $d=6,7$ and CG. Figure~\ref{fig:nsC2v}(a) shows $f_{X'_2}(x)$ against $x$, and the good data collapse suggests that $C_2 \sim L^{1+d/2}$, following the GFP asymptotics. Figure~\ref{fig:nsC2v}(b) implies $C_2 \sim V^{2/3}$ for $d=6,7$ and CG. The data from 6D and 7D collapse well, but there is a little discrepancy with CG, which may originate from the choice of constant $a_0$.  
The different scaling behaviors of the vanishing sectors imply $6$ is a special dimension, which gives another hint for geometric upper critical dimension $d_p=6$. \par

\subsubsection{The Ising sector}
\label{Ising sector}
\begin{figure}[t]
    \centering
    \includegraphics[width=0.50\textwidth]{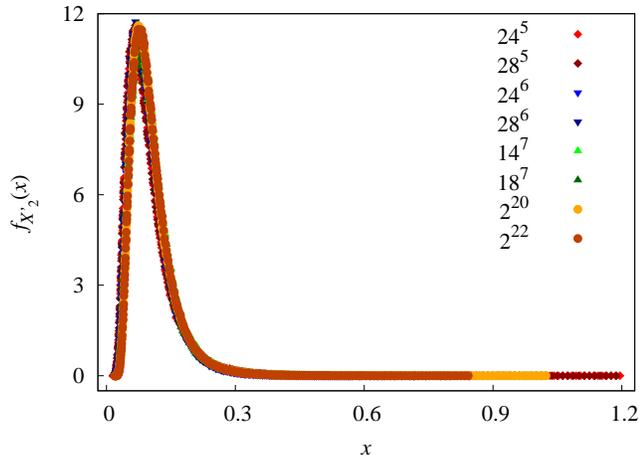}
    \caption{The probability density function of the rescaled second-largest cluster in the Ising sector. The variable $X_2' := \scrC_2 \ln L /(aL^{1+d/2})$ with $a=4.70, 4.22, 4.67$ for $d=5,6,7$, respectively and $ X_2' := \scrC_2/(V^{1/2} \ln V)$ for CG.}
    \label{fig:ns2C2}
\end{figure}

Except for the vanishing sector, we also define the Ising sector, whose probability approaches 1 as system volume goes to infinity. 
To suffer from less finite-size corrections, we include only the configurations with $\scrC_1 \ge V^{3/4}$ to the Ising sector. In this sector, its largest cluster follows the CG asymptotics $C_1 \sim V^{3/4}$. For the second-largest cluster $\scrC_2$, we define $X_2'=\scrC_2 \ln L /(aL^{1+d/2})$ for high-d tori and $X_2'=\scrC_2/(V^{1/2}\ln V)$ for CG, and  the data can collapse well  for various systems, as shown in Fig.~\ref{fig:ns2C2}.\par

\subsection{Existence of the two-length-scale behavior}
\label{two scales}

\subsubsection{The finite-size fractal dimensions  $d_\textsc{l1}$ and $d_\textsc{l2}$}
\label{finite-size fractal dimension}
In this section, we study the two-length-scale behavior by extracting the 
finite-size fractal dimensions $d_\textsc{l1}$ and $d_\textsc{l2}$.
We first recall the theoretical study on the CG. In Ref.~\cite{Luczak2006Phase}, 
it was numerically observed that the critical FK Ising model on the CG has two-length-scale behavior, 
in which the largest cluster $C_1 \sim V^{3/4}$ and the second-largest cluster $C_2\sim {\cal O}(V^{1/2} \ln V)$ 
have different scaling behaviors, which was further numerically testified in Ref.~\cite{Fang2021Percolation}. 
Later, numerical results on the 5D FK Ising model also showed its two-length-scale behavior \cite{Fang2020Complete}.

\begin{table}[b]
\centering
\begin{tabular}{|l|llllll|}
\hline 
$d$  &$L_{\rm m  }$  	&$d_\textsc{l1}$ 	&$a_0$ 	&$b_1$ 	 	&$y_1$  	& $\chi^2/{\rm DF}$ 	\\
\hline 
        &6     &3.743(2)  	&1.160(7)  	&-1.1(3)            &-1.9(2)   	&4.4/6\\ 
$5$     &8     &3.743(3)  	&1.16(1)   	&-1(1)              &-2.1(6)   	&4.3/5\\ 
\hline
    &6     &4.508(4)  	&1.00(1)   	&0.5(3)    	&-2     &1.7/5\\ 
    &8     &4.510(7)  	&1.00(2)   	&0.8(9)    	&-2     &1.6/4\\ 
$6$ &6     &4.506(3)  	&1.009(9)  	&3(2)      	&-3     &1.8/5\\ 
    &8     &4.509(6)  	&1.00(2)   	&8(9)      	&-3    	&1.5/4\\ 
    &10    &4.505(9)  	&1.01(3)   	&-5(26) 	&-3    	&1.3/3\\ 
\hline 
    &5     &5.196(5)  	&1.12(2)   	&-12(2)    		&-3	    &6.4/6\\ 
    &6     &5.188(8)  	&1.14(3)   	&-17(5)    	    &-3 	&4.9/5\\ 
$7$ &7     &5.18(1)   	&1.18(5)   	&-28(13)		&-3 	&4.2/4\\ 
    &4     &5.190(4)  	&1.14(1)   	&-2.0(1)   	   	&-2   	&6.6/7\\ 
    &5     &5.183(7)  	&1.16(2)   	&-2.5(4)   	 	&-2  	&5.0/6\\ 
    &6     &5.18(1)   	&1.19(4)   	&-3.1(9)   	 	&-2	    &4.3/5\\ 
\hline 
\end{tabular} 
\caption{Estimates of the finite-size fractal dimension $d_\textsc{l1}$ with $d \ge 5$ via  ansatz Eq.~\eqref{eq:fitting_ansatz1}. The conjectured values of $d_\textsc{l1}$ are 15/4, 9/2, 21/4 for $d=5,6,7$, respectively.}
\label{tab:fits_567DC1L} 
\end{table} 

\begin{figure}[t]
    \centering
    \includegraphics[width=0.5\textwidth]{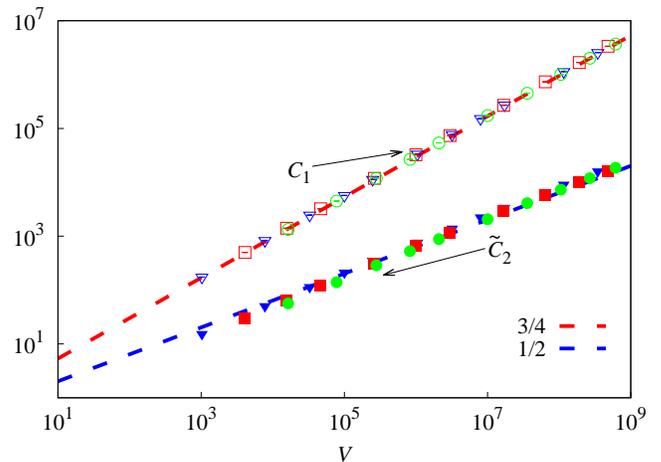}
    \caption{The FSS behaviors of the largest cluster $C_1$ (hollow points) and second-largest cluster $\tilde{C_2} := C_2 \ln L/L$ (solid points) for various system sizes with $d=5$ (blue), $d=6$ (red) and $d=7$ (green). These scaling behaviors follow the CG asymptotics. }
    \label{fig:C12_DL}
\end{figure}
   
\par 

In a previous study \cite{Fang2020Complete}, we argued that the scaling behaviors of the FK Ising model on the 5D tori  
are simultaneously governed by the CG asymptotics and the GFP asymptotics,
which were supported by large-scale Monte Carlo simulation results. 
Based on it, we argue that the largest cluster $C_1 \sim L^{3d/4}$ and 
the second-largest cluster $C_2 \sim L^{1+d/2}/\ln L$, 
which gradually converges to the CG scaling behavior $V^{1/2} \ln V$ in the $d \to \infty$.
Note that the appearance of the multiplicative logarithmic correction $1/\ln L$
in the scaling of $C_2$ is conjectured purely according to our numerical data, see below for details. 
Figure~\ref{fig:C12_DL} plots $C_1$ and $\tilde{C}_2 \equiv C_2\ln L/L$ versus the system volume $V$, 
and data from various systems collapse well with slopes consistent with $3/4$ and $1/2$, respectively. 
To extract the value of the finite-size fractal dimensions $d_\textsc{l1}$ and $d_\textsc{l2}$, 
we perform the least-square fits to the MC data.

\begin{table}[b]
\centering
\setlength\tabcolsep{1.0pt}
\begin{tabular}{|l|lllllll|}
\hline 
$d$ &$L_{\rm m}$  	&$d_\textsc{l2}$ 	&$\hat{y}_{\scrO}$ &$a_0$ 	&$b_1$ 	 	&$y_1$ 	& $\chi^2/{\rm DF}$ 	\\
\hline 
    &10    &7/2    	&-1.05(1)   	&1.09(2)   	&-1.25(4) 	       &-1/2   	 &1.9/5\\ 
    &12    &7/2    	&-1.06(2)   	&1.11(3)   	&-1.29(6)          &-1/2       &1.2/4\\ 
    &16    &7/2    	&-1.08(3)   	&1.15(6)   	&-1.4(1)  	       &-1/2   	 &0.6/3\\ 
$5$ &10    &3.490(2) &-1         	&1.06(1)   	&-1.19(2)  	      &-1/2       &1.5/5\\ 
    &12    &3.489(3) &-1          	&1.07(1)   	&-1.20(3)         &-1/2       &1.1/4\\ 
    &16    &3.485(6) &-1          	&1.09(3)   	&-1.25(7)  	      &-1/2       &0.7/3\\ 
\hline 
    &8     &4        	&-0.99(3)   &0.94(3)   	&-17(2)    		&-2   	&7.7/4\\ 
    &10    &4        	&-1.07(5)   &1.05(6)   	&-26(5)      	&-2   	&2.9/3\\ 
    &12    &4           &-1.3(2) 	&1.3(3)    	&-59(33)  		&-2   	&1.5/2\\ 
$6$ &8     &4.003(7)   	&-1     	&0.94(2)   	&-17(1)         &-2 	&7.8/4\\ 
    &10    &3.98(1)    	&-1     	&1.02(4)   	&-25(4)    	    &-2 	&2.9/3\\ 
    &12    &3.94(4)   	&-1      	&1.2(2)    	&-47(22)    	&-2    	&1.5/2\\ 
\hline
    &7     &4.49(1)   &-1  	    &1.64(7)    &-1.9(1)       	&-1/2  	  	&2.7/5\\ 
    &8     &4.48(2)   &-1  	    &1.7(1)     &-2.0(2)   	    &-1/2  	  	&2.5/4\\ 
$7$ &7     &9/2      &-1.04(4) 	&1.7(1)  	&-2.0(2)   	    &-1/2    	&2.7/5\\ 
    &8     &9/2    	&-1.06(7)  	&1.7(2)    	&-2.1(3)   	    &-1/2    	&2.4/4\\
\hline 
\end{tabular} 
\caption{Estimates of the finite-size fractal dimension $d_\textsc{l2}$ for $d\ge5$ with multiplicative logarithmic corrections via the ansatz Eq.~\eqref{eq:fitting_ansatz2}. The conjectured values of $d_\textsc{l2}$ are 7/2, 4, 9/2 for $d=5,6,7$, respectively. }
\label{tab:fit_567DC2L1} 
\end{table} 

We first consider the largest cluster $C_1$ for $d>4$.  We perform the least-square fits to it 
via the  ansatz Eq.~\eqref{eq:fitting_ansatz1}. The fitting results are summarized 
in Table~\ref{tab:fits_567DC1L}. For $d=5$, we first set $b_2=c_0=0$ and leave other parameters free, 
and we obtain $d_\textsc{l1} = 3.743(2)$ and $y_1 =-1.9(2)$  with $\Lm = 8$.  
Leaving $c_0$ free also gives a consistent estimate $d_\textsc{l1} = 3.736(11)$. 
By comparing with various ansatz, we finally get the estimate $d_\textsc{l1}=3.74(2)$, 
which is consistent with the expected value $15/4$. 
For $d=6$, we first set $b_2=0$ and leave other parameters free, but it does not yield stable results. Consequently, we fix the correction exponent $y_1$ and obtain the estimates $d_\textsc{l1}=4.510(7)$ for $y_1=-2$ and $d_\textsc{l1}=4.505(10)$ for $y_1=-3$.
By comparing various ansatz, we obtain the final estimate $d_\textsc{l1}=4.51(1)$,
 which is consistent with the expected value $9/2$. 
For $d=7$, following a similar procedure, we obtain the final estimate $d_\textsc{l1} = 5.18(2)$, 
which is close to the expected value $21/4$. 
The discrepancy between the estimate and the expected value may be due to the fact that the precision of 
the critical threshold is not high enough such that the true critical point is slightly away 
from the quoted value in Ref.~\cite{Lundow2015Discontinuity}. \par 

\begin{figure}[t]
    \centering
    \includegraphics[width=0.52\textwidth]{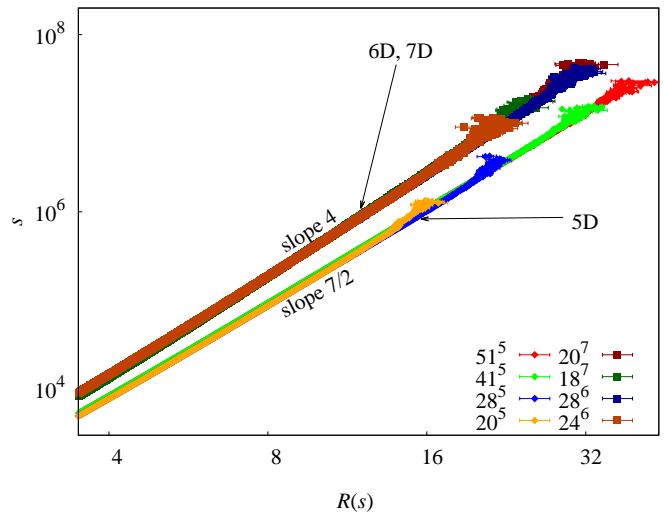}
    \caption{The log-log plot of the clusters size $s$ versus their radius $R(s)$ for $d\ge5$. 
    The slopes display the thermodynamic fractal dimension $D_\textsc{f}$.} 
    \label{fig:C2_DF}
\end{figure}

\begin{table}[b]
\centering
\setlength\tabcolsep{1.0pt}
\begin{tabular}{|l|l|lllllll|}
\hline 
$d$& $\scrO$ &$L_{\rm m}$  	&$y_{\scrO}$ 	&$a_0$ & $b_1$ 	& $c_0$    &$y_1$ 	& $\chi^2/{\rm DF}$ 	\\
\hline 
&$C_1$  &6        &3.768(4)     &2.25(3)    &3(2)       &\,7(9)       &-1.8(3)   &2.0/5  \\
$5$&    &8        &3.767(6)  	&2.27(5)  	&6(11)    	&-10(58)  	&-2(1)     &1.6/4\\ 

\cline{2-9} 
&$C_2$  &6     &3.620(3)  	&2.38(3)  	&6(2) 	    &\,3(4)     &-2.3(2)        &1.7/5\\ 
&       &8     &3.616(4)  	   &2.40(3)   	&27(50)   	&-35(65)     &-2.9(7)         &1.0/4\\ 
\hline 

&       &6        &4.56(6)   &0.36(9)    &3.0(1)         &-          &-1.05(3)   &6.1/4  \\
&$C_1$  &4        &4.59(3)  	&0.30(3)  	&2.71(1)    	&24(1)  	&-1   	   	&6.5/6\\ 
$6$ &       &5        &4.58(3)   &0.32(4)  	&2.707(8)   	&26(2)    	&-1   	   	&5.4/5\\ 
  
 \cline{2-9}  
& $C_2$  &4     &3.95(2)  	&1.58(9)  	&3.9(5) 	&3(2)     &-1.7(2)        &2.5/5\\ 
&       &5     &4.03(5)  	&1.2(3)   	&2.5(2)   	&-10(2)     &-1.0(2)         &1.0/4\\ 
\hline 
    &       &6        &4.58(4)   &0.6(1)    &3.1(6)         &-          &-1.3(2)   &2.9/5  \\
$7$& $C_1$  &7        &4.55(6)  	&0.7(2)  	&4(3)    	&-  	&-1.5(5)   	   	&2.7/4\\ 
 \cline{2-9}  
     &      &5        &4.08(3)   &1.4(1)    &3.8(6)         &-          &-1.6(2)   &5.1/6  \\
     &  $C_2$         &6        &4.06(4)  	&1.5(2)  	&5(2)    	&-  	&-1.8(2)   	   	&4.9/5\\
\hline 
\end{tabular} 
\caption{Estimates of the thermodynamic fractal dimensions $d_\textsc{f1}$ and $d_\textsc{f2}$ with $d=5,6,7$. The exponent $y_{\scrO}$ corresponds to $d_\textsc{f1}$ for $C_1$ and $d_\textsc{f2}$ for $C_2$ in each dimension.}
\label{tab:fits_567DC12R} 
\end{table} 

We then consider the second-largest cluster $C_2$. We assume that it scales as $C_2 \sim L^{1+d/2}/\ln L$ for $d\ge5$. 
We perform the least-squares fits to it via the ansatz Eq.~\eqref{eq:fitting_ansatz2}. 
Taking $d=5$ as an example, leaving $y_{\scrO}$ and $\hat{y}_{\scrO}$ free can not yield reasonable results. 
We then fix $y_{\scrO}=7/2$ and obtain the estimate $\hat{y}_{\scrO} = -1.07(6)$. 
We then fix $\hat{y}_{\scrO}=-1$ and obtain $y_{\scrO} = 3.49(1)$, which is consistent with 
the expected value $7/2$ .
Following a similar procedure, we obtain the estimates $d_\textsc{l2}= 3.95(7), 4.48(3)$ for $d=6,7$,
respectively, and the logarithmic correction exponents are consistent with $-1$. 
These estimates are consistent with our conjecture $C_2 \sim L^{1+d/2}/\ln L$, 
and the fitting results are summarized in Table~\ref{tab:fit_567DC2L1}. \par

In addition, we also try to fit the $C_2$ data to the ansatz Eq.~\eqref{eq:fitting_ansatz1} without logarithmic corrections.
We obtain the estimates $d_\textsc{l2}= 3.32(1), 3.71(3), 4.32(2)$ for $d=5,6,7$, respectively. These estimates all deviate away from the expected values. This is why we believe there exists the logarithmic correction $1/\ln L$ in the scaling behavior of $C_2$.

\begin{figure}[t]
    \centering
     \includegraphics[width=0.55\textwidth]{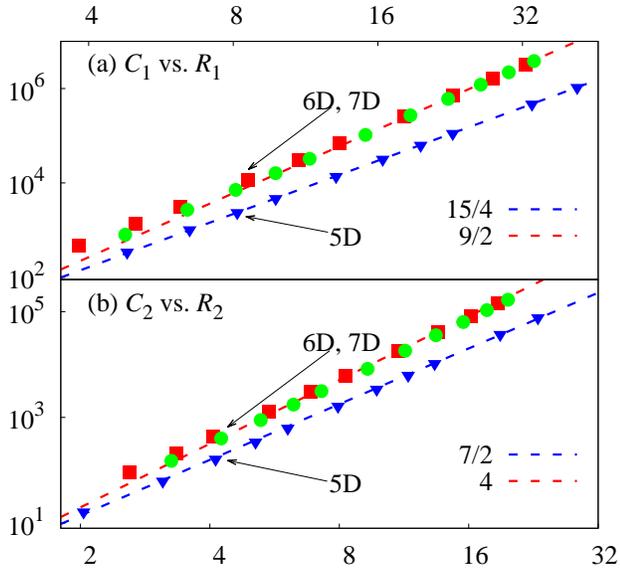}
    \caption{The log-log plot of (a) the largest cluster $C_1$ versus its radius $R_1$  and (b) the second-largest cluster $C_2$ versus its radius $R_2$ for various system sizes in $d=5$ (blue), $d=6$ (red) and $d=7$ (green).}
    \label{fig:C12_DF}
\end{figure}

\subsubsection{The thermodynamic fractal dimensions $d_\textsc{f1}$ and $d_\textsc{f2}$} 
\label{Thermodynamic fractal dimension}

\begin{figure}[t]
    \centering
    \includegraphics[width=0.52\textwidth]{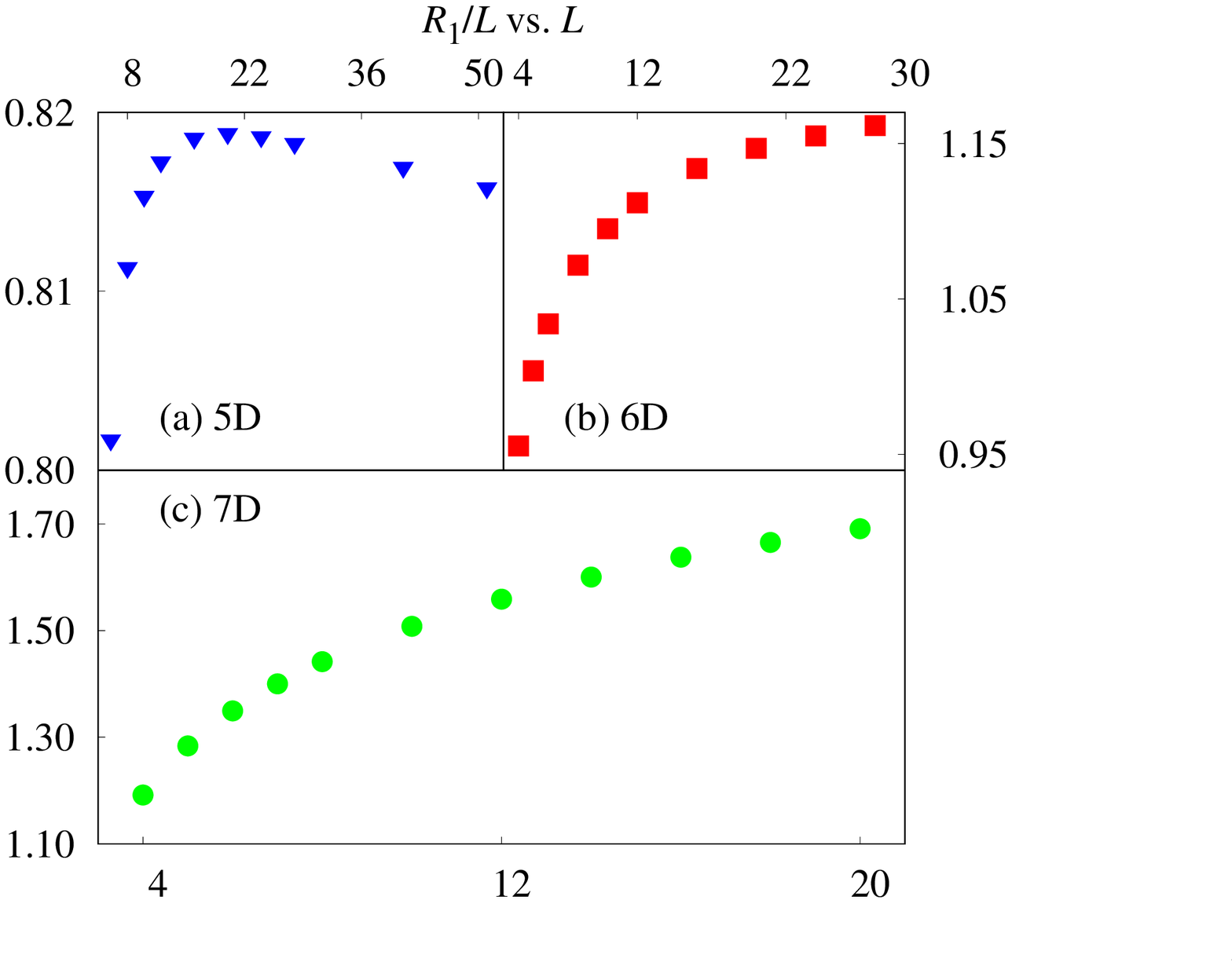}
        \caption{The FSS  behavior of the rescaled  radius $R_1/L$ versus system size $L$ for (a) $d=5$, (b) $d=6$ and (c) $d=7$. For $d < d_p$, the radius $R_1$ is bounded by the linear system size $L$, while it increases faster than $L$ for $d>d_p$.}
    \label{fig:C1ra}
\end{figure}

We then consider the thermodynamic fractal dimensions $d_\textsc{f1}$ and $d_\textsc{f2}$. To extract their values, we plot the largest cluster $C_1$ versus its radius $R_1$ and the second-largest cluster $C_2$ versus its radius $R_2$ in the log-log scale, as seen in  Fig.~\ref{fig:C12_DF}. The slopes of  lines indicate the fractal dimensions. We find that $d_\textsc{f1}$ is consistent with the finite-size fractal dimension $d_\textsc{l1}=3d/4$ for $4 < d < 6$,  while for $d\ge6$, it is consistent with $9/2$, independent of the spatial dimension. 
For the fractal dimension $d_\textsc{f2}$, we find that it is consistent with $1+d/2$ for $4 < d <6$ and $4$ for $d \ge 6$. \par  

We then perform the least-square fits to the MC data. 
For $d\ge5$, we assume $C_1$ and $C_2$ suffer free from logarithmic correction, and we take the fitting ansatz Eq.~\eqref{eq:fitting_ansatz1}. The results are summarized in  Table~\ref{tab:fits_567DC12R}. Considering the systematic error from various fitting ansatz, we finally estimate $d_\textsc{f1} = 3.76(1), 4.6(1), 4.55(12)$ and $d_\textsc{f2}= 3.61(3), 4.0(1), 4.1(1)$ for $d=5,6,7$, respectively. 
Except for $d_\textsc{f2}$ in $d=5$, all of these estimates are consistent with our conjecture. The slight disagreement of the estimate of $d_\textsc{f2}$ to its expected value at 5D might be due to the potential logarithmic corrections. 

We next study the thermodynamic fractal dimension for clusters other than $C_1$ and $C_2$. We plot the size of these clusters $s$ versus their radius $R$ for $d\ge 5$ in Fig.~\ref{fig:C2_DF}. In the log-log plot, the slopes of lines indicate the value of $d_\textsc{f2}$, 
and we find for all critical clusters with medium size that the fractal dimension is consistent with $d_\textsc{f2}$.  
As Fig.~\ref{fig:C2_DF} shows, we find for $d=5$ that the slope of the line is consistent with $1+d/2$, 
following the GFP asymptotics, while for $d \ge d_p$, 
it has a good data collapse with a slope consistent with $4$, 
which has the same value as the high-d percolation model. 

To summarize, for $d>4$, the two-length-scale behavior begins to appear with the finite-size fractal dimensions $d_\textsc{l1}=3d/4$ and $d_\textsc{l2}=1+d/2$, consistent with the CG asymptotics. For $d \ge 6$, the thermodynamic fractal dimensions $d_\textsc{f1}$,$d_\textsc{f2}$ are no longer the same as the finite-size fractal dimensions and consistent with dimension-independent constants 9/2 and 4, respectively. The two-length-scale behavior of the fractal dimensions gives solid support for the simultaneous existence of the two upper critical dimensions.

\begin{figure}[t]
    \centering
    \includegraphics[width=0.52\textwidth]{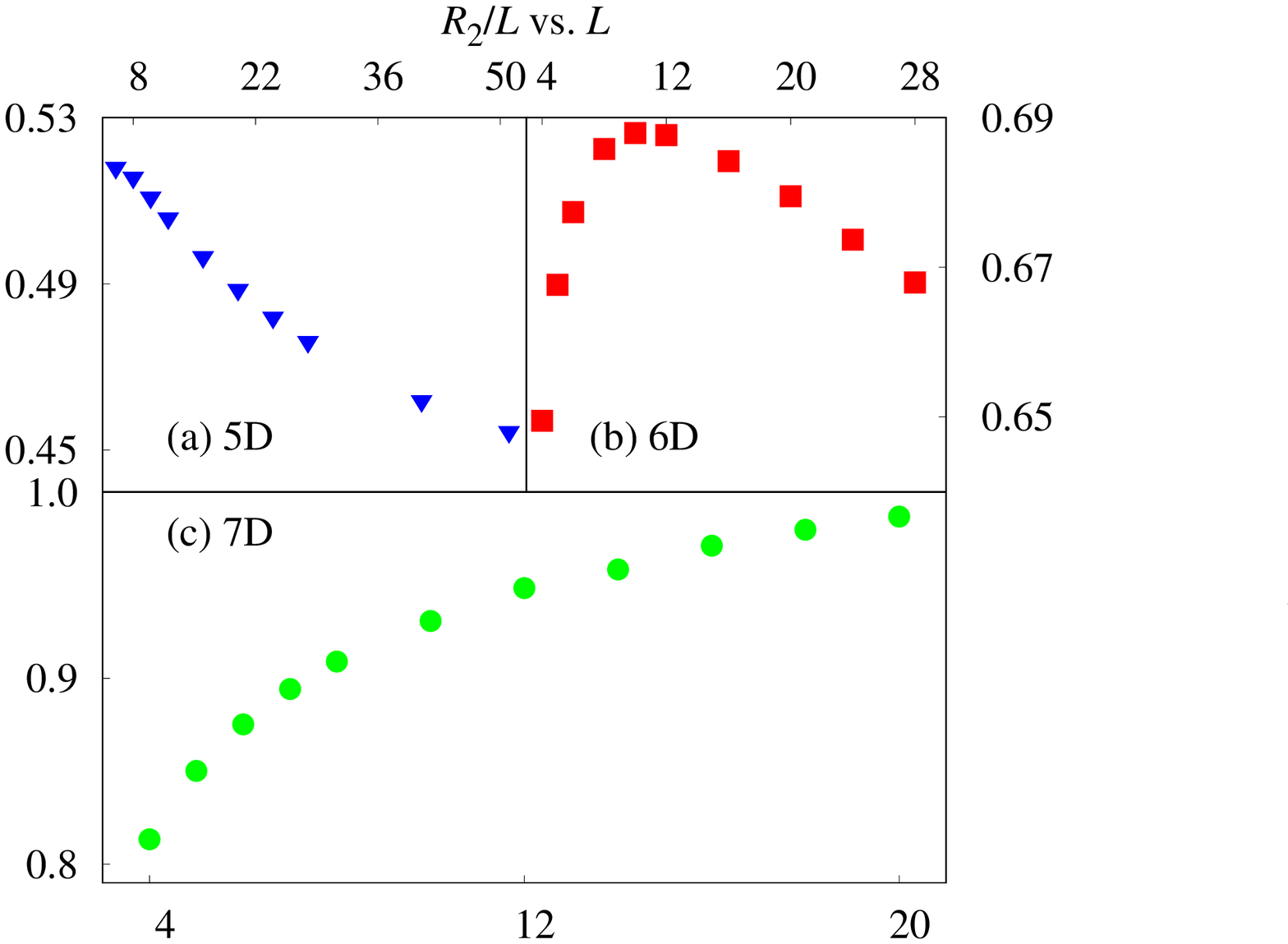}
    \caption{The FSS behavior of the rescaled  radius $R_2/L$ versus system size $L$ for (a) $d=5$, (b) $d=6$ and (c) $d=7$. For $d \le d_p =6$, the radius $R_2$ is bounded by the linear system size $L$, while it increases faster than $L$ for $d>d_p$.}
    \label{fig:C2ra}
\end{figure}

\begin{table}[b]
\centering
\begin{tabular}{|l|llllllll|}
\hline 
$d$ &$L_{\rm m}$  	&$d_\textsc{r1}$ 	&$a_0$ & $b_1$ 	& $b_2$   &$y_1$ 	&$y_2$ 	& $\chi^2/{\rm DF}$ 	\\
\hline 
5      &8     &0.999(2)  	&0.809(9)  	&0.17(3)   	&-0.61(5)  	&-2/3   &-4/3    &2.6/5\\ 
       &10    &0.997(3)  	&0.82(1)   	&0.11(6)   	&-0.5(1) 	&-2/3   &-4/3    &1.5/4\\ 
\hline 
6 &10    &1.001(2)  	&1.19(1)   	&-1.01(4)  	&-   	&-1  	&-    	&4.0/3\\ 
  &12    &0.997(4)  	&1.21(2)   	&-1.10(7)  	&-   	&-1   	&-    	&1.8/2\\

\hline 
&6     &1.102(7)  	&1.30(4)   	&-1.2(2)   	&0.5(5)    	&-1        	&-2    	&6.7/5\\ 
&7     &1.09(1)   	&1.35(7)   	&-1.4(5)   	&1(1)      	&-1        	&-2    	&6.2/4\\ 
7&6     &1.08(2)   	&1.44(10)  	&-0.9(3)   	&-0.2(4)   	&-2/3    	&-4/3   &6.5/5\\ 
 &7     &1.07(3)   	&1.5(2)    	&-1.3(7)   	&0.4(9)    	&-2/3    	&-4/3   &6.0/4\\
\hline 
\end{tabular} 
\caption{The fitting results of the radius of the largest cluster $R_1$ for $d=5,6,7$. For $d \le 6$, it is of order $L$, while it deviates from $L$ for $d> 6$.}
\label{tab:C1_ra} 
\end{table} 

\subsubsection{The scaling behavior of radius $R_1$ and $R_2$}
\label{The scaling behavior of the R1 and R2}
From the scaling behaviors $C_1 \sim L^{d_\textsc{l1}} \sim R_1^{d_\textsc{f1}}$ 
and $C_2 \sim L^{d_\textsc{l2}} \sim R_2^{d_\textsc{f2}}$, 
we have $R_1 \sim L^{d_\textsc{l  1}/d_\textsc{f1}}$ and $R_2 \sim L^{d_\textsc{l2}/d_\textsc{f2}}$.
Thus, the scaling behaviors of $R_1$ and $R_2$ follow 
\begin{align}
R_1 \sim R_2 \sim L,  \hspace{24mm} & d \le 6, \nonumber  \\
R_1 \sim L^{d/6},  \hspace{5mm} R_2 \sim L^{1/4+d/8}, & d > 6. 
\end{align}
Note that we ignore the logarithmic corrections for $R_2$. 

We first consider $R_1$ and plot  $R_1/L$ versus $L$ for various systems, as shown in Fig.~\ref{fig:C1ra}.
For $4 < d < 6$, as system size increases, we find that $R_1/L$ converges to a constant for $4 < d < 6$ but increases for $d \ge 6$.
This means the largest cluster does not wind around the torus below 6D but winds extensively above 6D.
To verify the precise scaling behavior of $R_1$, we perform the least-square fits to MC data 
via the ansatz Eq.~\eqref{eq:fitting_ansatz1}, where $\scrO$ corresponds to $R_1$ and $y_{\scrO}$ 
corresponds to $d_\textsc{r1}$. The fitting results are summarized in Table~\ref{tab:C1_ra}. 
We find for $d\le 6$ that $d_\textsc{r1}$ is consistent with 1, while it is larger than 1 for $d=7$. 
Nevertheless, it is not consistent with $7/6$, which may be due to the fact that 
the precision of the critical threshold is not high enough.\par  

We then consider $R_2$. In Fig.~\ref{fig:C2ra}, we plot the rescaled radius $R_2/L$ versus its system size $L$. As $L$ increases, $R_2/L$ converges to a constant for $d \le 6$, but it increases for $d=7$. 
\par

\begin{figure}[t]
    \centering
    \includegraphics[width=0.52\textwidth]{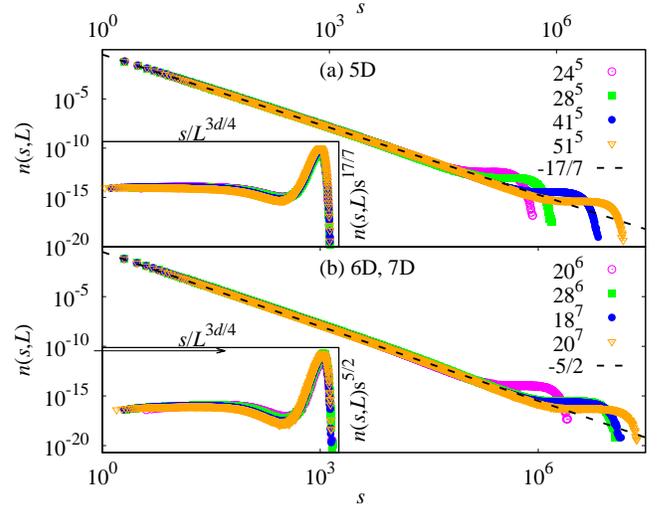}
    \caption{The cluster-number density $n(s,L)$ for (a) $d=5$ and (b) $d=6,7$. The Fisher exponent $\tau$ is consistent with $1+\frac{d}{1+d/2}$ for $d<d_p$ and $5/2$ for $d \ge d_p$.}
    \label{fig:nsd567}
\end{figure}

\subsubsection{The cluster number density $n(s,L)$}
\label{cluster number density}

We then consider the cluster-number density $n(s,L)$. Generally, it is expected that
\begin{equation}
n(s,L) \sim s^{-\tau} \tilde{n}(s/L^{d_\textsc{l1}}) \; ,
\label{eq:ns_FSS}
\end{equation}
where $\tau$ is the Fisher exponent. The universal scaling function $\tilde{n}(x)$ 
is approximately a constant for $x \ll 1$ and drops quickly for $x \gg 1$. The exponents $\tau$ and $d_\textsc{l1}$ are not independent but obey the hyperscaling relation 
\begin{equation}
\label{eq:fisher_HR}
\tau = 1 + d/d_\textsc{l1},
\end{equation}
The above scaling behavior of $n(s,L)$ has been observed for percolation models in various dimensions and random-cluster models below $d_c = 4$~\cite{Hou2019Geometric}.\par

We then plot $n(s,L)$ versus cluster size $s$ for $5\leq d \leq 7$, shown in Fig.~\ref{fig:nsd567}. 
As we can see, it first displays a power-law behavior with the slope being the Fisher exponent $\tau$, then it enters a plateau, and finally it decays significantly. Our data show that $\tau$ is consistent with $17/7$ for $d=5$ and $5/2$ for $d=6,7$; the latter was conjectured in Ref.~\cite{Chayes1999Meanfield} and is consistent with the value of the FK Ising model 
on the CG~\cite{Fang2021Percolation} and the percolation model on high-d tori and the CG ~\cite{Huang2018Critical}. Thus, it supports $d_p = 6$ is an upper critical dimension.

We note that, using the values of $d_\textsc{l1}$ in Tab.~\ref{tab:estimate_DFDL}, the scaling relation Eq.~\eqref{eq:fisher_HR} is broken. Using these values of $\tau$, one can obtain an effective fractal dimension $d_{\rm eff}$ using Eq.~\eqref{eq:fisher_HR}. For $d=5$, since $\tau = 17/7$, one has $d_{\rm eff} = 7/2$, consistent with $d_\textsc{l2}=1+d/2$ from the GFP prediction. For $d\ge6$, using $\tau = 5/2$, one has $d_{\rm eff} = 2d/3$ which is the fractal dimension of the percolation clusters on high-d tori and CG.

\subsubsection{The number of spanning cluster $N_s$}
\label{number of spanning cluster} 

\begin{figure}[t]
\centering
\includegraphics[width=0.52\textwidth]{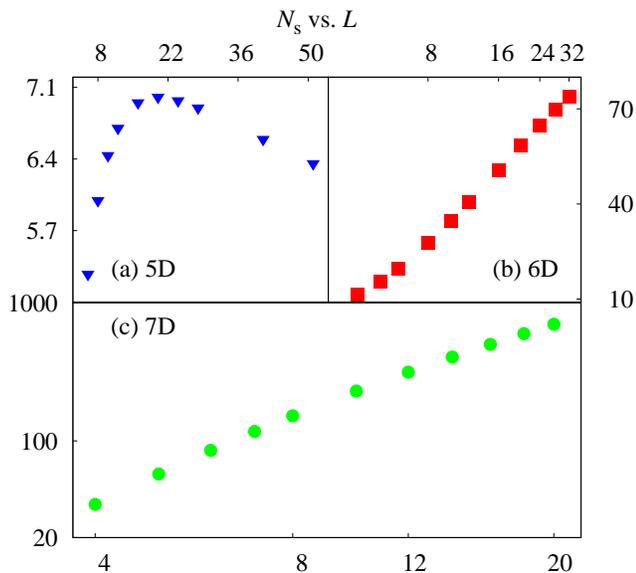}
\caption{The number of spanning clusters $N_s$, which is bounded for (a) $d=5$ and 
diverges for $d=6$ and $d=7$. The semi-log plot in (b) implies $N_s \sim \ln L$ for $d=6$. 
The log-log plot in (c) implies $N_s$ for $d=7$ diverges as a power law.} 
\label{fig:Nspan}
\end{figure}

We next study the number of spanning clusters $N_s$. Recall that a cluster 
is spanning if its unwrapped distance ${\cal U}$ is not less than the linear size $L$.
We plot $N_s$ versus $L$ for $d=5,6,7$ in Fig.~\ref{fig:Nspan}. For $d<6$, we see $N_s$ converges to a bounded value, while for $d \ge 6$, $N_s$ increases as system size increases.
For $d=6$, the straight line in the semi-log plot in Fig.~\ref{fig:Nspan}(b) suggests $N_s \sim \ln L$, and the straight line in the log-log plot with a slope close to 1 in Fig.~\ref{fig:Nspan}(c) suggests $N_s \sim L$ for $d=7$, which is consistent with the observation on the high-d percolation~\cite{Lu2023Finite}.
\par  

The divergence of $N_s$ above 6D can be understood from the behavior of $n(s,L)$.
As discussed in Sec.~\ref{Thermodynamic fractal dimension}, for clusters except the largest one, their sizes scale with the radius of gyration as $s\sim R^4$ for $d\geq 6$. It is reasonable to expect the unwrapped distance ${\cal U}$ of a cluster is of the same order as $R$. Thus, it follows that a spanning cluster has $R$ no less than $L$, and thus its size $s$ is at least of order $L^4$. Thus, the number of spanning clusters above 6D can be calculated as
\begin{align}
\label{eq:ns_highd}
N_s  \sim L^d \int_{L^4}^{L^d} n(s, L) \diff s \sim L^{d-6}. \nonumber
\end{align}

So $N_s$ diverges as $L^{d-6}$ for $d>6$, which is the same as the percolation model on high-d tori~\cite{Aizenman1997number}. In the marginal case $d = 6$, possibly $N_s$ diverges logarithmically.

\begin{table}[b]
    \centering
    \begin{tabular}{|l|lllllll|}
    \hline 
    $d$ &$L_{\rm m}$  	&$y_{\scrO}$ 	&$a_0$ 	&$b_1$ 	&$c_0$ 	&$y_1$ 		& $\chi^2/{\rm DF}$ 	\\
    \hline 
    $7$ &7     &0.97(1)   	&90(6)     	&-292(27)	&250(27)	&-1/2    	&1.5/4\\ 
        &8     &0.99(3)   	&80(11) 	&-244(52)	&201(53)	&-1/2    	&0.5/3\\ 
    \hline 
    \end{tabular} 
    \caption{The fitting results of the number of spanning cluster $N_s$ for $d=7$, which is consistent with the conjecture $N_s \sim L$.}
    \label{tab:Nspan} 
    \end{table} 
    
To verify the scaling of $N_s$, we perform the least-squares fits to the MC data. 
For $d=6$, we use the logarithmic fitting ansatz Eq.~\eqref{eq:fitting_ansatz2} with $y_{\scrO} = 0$.
We first leave $\hat{y}_{\scrO}$ free, but no stable results can be obtained. We then fix $\hat{y}_{\scrO}=1$, $y_1=-1$ and leave $a_0, b_1, c_0$ free, and we obtain stable fits when $\Lm = 16$, which gives $a_0=19(2)$, $b_1=-150(20)$ and $c_0=23(9)$ with the residuals $\chi^2\approx 0.9$. 
For $d=7$, we fit the MC data to the ansatz Eq.~\eqref{eq:fitting_ansatz1}. 
We first set $b_1=b_2=0$ and leave $a_0$, $c_0$ and $y_{\scrO}$ free, which gives unstable results. Leaving the correction exponent $y_1$ or $y_2$ free cannot yield stable results. Thus, we fix the exponent $y_1$ to various values, and we leave $a_0$, $b_1$, $c_0$ and $y_{\scrO}$ free.  The fitting results are summarized in Table~\ref{tab:Nspan}. 
\par 

\section{Results near criticality}
\label{results2}
In this section, we consider the critical behavior away from the critical point $K_c$. 
We study the coupling strength $K = K_c - aL^{-\lambda}$ with $\lambda > 0$, i.e., the reduced temperature $t = \frac{1}{K_c} a L^{-\lambda}$. 
When $a>0$, the critical point $t=0$ is approached from the high-T side, and $a<0$ is from the low-T side.  
We study the scaling behavior for different values of $\lambda$. 
For $d >d_c$, we find there are asymmetric behaviors as $K_c$ is approached from different sides, and the asymmetric behaviors are different for $4<d<6$ and $d\ge 6$. 
On the high-T side, we find that there exists more than one scaling window.

\subsection{High-temperature side}
\label{high-T asymptotics}   
We recall some scaling behaviors that are believed to hold near criticality. 
First, for $d \geq 4$, the leading FSS behavior of the magnetic susceptibility $\chi$ reads 
\begin{equation}
\chi(t,L) \simeq L^{2y_h^* - d} \tilde{\chi} (t L^{y_t^*}) = L^{d/2} \tilde{\chi} (t L^{d/2}) \; , 
\label{eq:chi_FSS}
\end{equation}
where the RG exponents $(y_h^*, y_t^*)=(3d/4, d/2)$ from the CG asymptotics have been used.
In the thermodynamic limit ($ L \! \to \! \infty$), 
$\chi$ exhibits singular scaling behavior as $\chi(t) \sim t^{-\gamma}$, 
and the mean-field value $\gamma=1$ can be obtained either from the GFP or the CG asymptotics.
We ask  how the FSS ansatz~(\ref{eq:chi_FSS}) transits to the thermodynamic scaling. 
Given any infinitesimal but finite $t$, the argument $x \equiv t L^{d/2}$ in the function 
$\tilde{\chi}(x)$ would diverge for sufficiently large $L$. 
To eliminate the explicit dependence of $L$, 
it is requested that $\tilde{\chi}(x) \! \sim \! x^{-1}$ for $x \! \gg \! 1$. 
Therefore, if the Ising critical point is approached from the high-T side as $t \! \sim \! L^{-\lambda}$ 
with $\lambda \! < \! d/2$, one has $\chi \! \sim \! L^{\lambda}$. 
Indeed, it was numerically observed~\cite{Zhou2018Randomlength} that $\chi$ on high-d tori follows 
\begin{equation}
\chi \sim  
    \begin{cases}
     L^{\lambda} &\text{if}~ \lambda<d/2\\ 
     L^{d/2}     &\text{if}~ \lambda \ge d/2\;.
     \end{cases}
\label{eq:chi_FSS_lambda}
\end{equation}
For $\lambda \geq d/2$,
since the renormalized scaling field $t L^{d/2}$ in Eq.~(\ref{eq:chi_FSS}) does not flow away,
the FSS behaviors of all the quantities, including $\chi$, 
should be the same as those at the critical point, as presented in Sec.~\ref{results1}.

Second,  for $0 \! < \! \lambda \! < \! d/2$, the cluster-number density $n(s,L)$ 
should obey a similar form to Eq.~(\ref{eq:ns_FSS}). 
Taking into account the potential two length scales, we separate the contribution from the largest cluster and write 
\begin{equation}
n(s,L) \sim s^{-\tau} \tilde{n}(s/s_\lambda) + L^{-d} f_{{\cal C}_1} (s, L) \; .
\label{eq:ns_FSS2}
\end{equation}
The factor $L^{-d}$ is for the density of the largest cluster,
which might have  a $\lambda$-dependent fractal dimension $d_{\lambda 1}$, and $f_{{\cal C}_1} (s,L)$ denotes
the cluster-size distribution of the largest cluster.
For $4 \! \leq \! d \! < \! 6$, the Fisher exponent is $\tau \! = \! 1+d/\df$ 
with $\df\! = \! 1+d/2$ from the GFP. 
Further, if the cut-off size  $s_\lambda \! \sim \! L^{d_\lambda}$, 
the number of clusters of size $\scrO(s_\lambda)$ 
would diverge as $N_\lambda \! \sim \! L^{d(1-d_\lambda/\df)}$ for $d_\lambda \! < \! \df$.
For $d \! \geq \! 6$, one has $\tau \! = \! 5/2$, and 
$N_\lambda$ is divergent for $d_\lambda \! <  2d/3$.

Since the magnetic susceptibility is identical to the second moment 
of cluster sizes as $\chi = \sum_{s} s^2 n(s,L)$, we have 
\begin{equation}
 \chi \sim (s_\lambda)^{3-\tau} + b L^{2 d_{\lambda1} - d} \; ,
 \label{eq:chi_FSS2}
\end{equation}
where $b$ is a positive constant. Depending on the value of $\lambda$, 
the leading scaling behavior of $\chi$ might come from the largest cluster,
from the remaining ones, or equally from both.

Third, as the temperature decreases, the correlation length $\xi$, 
corresponding to the diameters of characteristic clusters, 
grows as $\xi  \! \sim \! t^{-\nu}$, with $\nu \! = \! \nu_g \! = \! 1/2$ 
from the GFP.
For $t \sim L^{-\lambda}$, one would have $\xi \sim L^{\lambda/2}$,
suggesting that $\lambda = \lambda_g = 2$ is a special value.
For $\lambda < \lambda_g$, the diameters of clusters are much smaller 
than the linear size $L$. 
For $\lambda > \lambda_g$, the correlation length $\xi$ might be restricted to 
be of order $L$ or it might increase faster than $L$. In this case, finite-size effects become important. 

Fourth, it is helpful to consider the two-point correlation function $g(r,L)$, 
which is the probability for two vertices with distance $r$ to be in the same cluster.
By definition, the susceptibility is $\chi \equiv \sum g(r,L)$,
where the translational invariance on the tori is used. 
For $\lambda < \lambda_g=2$, we expect $g(r) \sim r^{2-d} \tilde{g}(r/\xi)$, 
where $\xi \ll  L$, and $\tilde{g}(x)$ drops exponentially for $x \! \gg \! 1$. 
For $\lambda > 2$, $g(r,L)$ develops a plateau for large distance
due to finite-size effects, which would contribute to the leading scaling behavior of $\chi$.
According to Eq.~(\ref{eq:chi_FSS_lambda}), one has  
\begin{equation}
g(r,L) \sim 
    \begin{cases}
         r^{2-d} \, \tilde{g}(r/L^{\lambda/2})  &\text{if}~~\lambda \in (0, \, 2) \\ 
         r^{2-d} + \scrO ( L^{\lambda-d}) &\text{if}~~\lambda \in [2, \, d/2) \\ 
         r^{2-d} + \scrO (L^{-d/2})       &\text{if}~~\lambda \in [d/2,\, \infty) \; .
    \end{cases}
\label{eq:g_FSS_lambda}
\end{equation}
Thus, for $\lambda > 2$, $g(r, L)$ exhibits the crossover behavior from the power-law decay, as predicted by GFP, to a distance-independent plateau.
The crossover happens at $r = \scrO \left(L^{(\lambda - d)/(2-d)}\right)$ for $2 < \lambda \leq d/2$ and at $r = \scrO(L^{d/2(d-2)})$ for $\lambda > d/2$.
The summation over the $r$-dependent part of the correlation function $g(r,L)$
gives $\chi \sim L^{\lambda}$ for $\lambda \leq 2$, and $\chi \sim L^2$ for $\lambda > 2$,
serving only as the subleading behavior of $\chi$.
To verify this scenario, one can define the magnetic structure factor as 
$\chi_{\bf k} \equiv \sum g({\bf r}) \exp (i {\bf k} \cdot {\bf r} )$ and take  the lowest momentum $|{\bf k}| = 2\pi/L$. 
Since the Fourier transformation would eliminate the contribution from the plateau of $g(r,L)$, 
one expects $\chi_{\bf k} \sim L^2$ for $\lambda \geq 2$, which was indeed observed 
in the previous studies~\cite{Wittmann2014Finitesize,Flores-Sola2016Role,Zhou2018Randomlength}. 

\subsubsection{$4 < d < 6$} 
For $\lambda < 2$, the critical point is approached at such a low speed that 
the diameters of the large clusters, though diverging, are much smaller than the linear size $L$. 
Finite-size effects are negligible and the critical behaviors of the
medium-size clusters are governed by the GFP. 
Namely, one has the cluster size $s \sim R^{d_g}$ with $d_g \equiv 1+d/2$ for
radii $ 1\! \ll \! R \! \ll \! L^{\lambda/2}$, and the cut-off size is 
$s_\lambda \sim L^{\lambda d_g /2}$. 
In Eq.~(\ref{eq:chi_FSS2}), the contribution to $\chi$ from the largest cluster 
is of order $\scrO(L^{\lambda -d(1-\lambda/2)}) < \scrO(L^{\lambda})$,  
and thus, the FSS of $\chi$ is from the summation over all the clusters. 
Actually, since the number $N_\lambda$ of large clusters of 
characteristic size $s_\lambda$ is $N_\lambda \sim L^{d(1-\lambda/2)}$, 
the contribution from these large clusters is already of order $\scrO (L^\lambda)$. 

For $\lambda=\lambda_g=2$, corresponding to the so-called Gaussian scaling 
window of width $\scrO(L^{-2})$, 
the correlation length $\xi$ reaches the order of $L$. 
There are only finite characteristic clusters $N_\lambda \sim \scrO(1)$;
a plateau of height $L^{2-d}$ develops in the scaling of the correlation function $g(r,L)$
for large $r$. 

For $\lambda \geq d/2$--i.e., the CG-Ising scaling window, 
the results in Sec.~\ref{results1} show that the unwrapped diameters of the largest 
and second-largest clusters are $R_1 \! \sim \! R_2 \! \sim \! L$.
Two lengthscales are exhibited in the sizes of clusters: 
the largest cluster scales as $C_1 \sim L^{3d/4}$, and all the other clusters 
have the Gaussian fractal dimension $d_g=1+d/2$.
The leading FSS $\chi \sim L^{d/2}$ is merely from the largest cluster, 
or, equivalently, from the plateau of $g(r,L)$.  

For $ \lambda_g \! < \! \lambda \! < \! d/2$, one naturally expects that 
$R_1 \! \sim \! R_2 \! \sim \! L$ and the Gaussian fractal dimension 
holds true for all clusters except $C_1$. 
Further, from the leading FSS behavior $\chi \sim L^{\lambda}$, 
one obtains $d_\textsc{f1}=d_\textsc{l1} = (d+\lambda)/2$ for the largest cluster. 

\begin{figure}[t]
    \centering
    \includegraphics[width=0.48\textwidth]{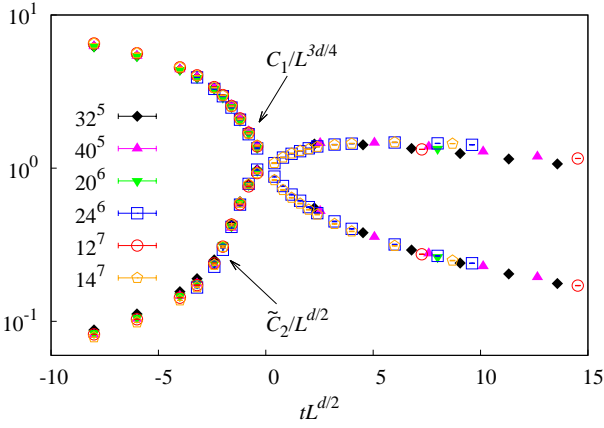}
    \caption{The CG-Ising scaling window illustrated by the rescaled cluster sizes 
    $C_1/L^{3d/4}$ and $\tilde{C}_2/L^{d/2} = C_2 \ln L/L^{1+d/2}$ on high-d tori with $d=5,6,7$. 
    Both $C_1$ and $C_2$ follow the CG asymptotics.}
    \label{fig:SWC12L12}
\end{figure}

\subsubsection{$d \geq 6$}

The scaling behaviors for $\lambda \leq 2$ should be similar to those 
for $ 4 \! \leq \! d \! < \! 6$, 
except that the Gaussian fractal dimension $\df=d_g$ should be replaced by 
$\df=4$, which can be regarded as being  from branching random walks. 
Namely, for medium-size clusters with radii $ 1 \! \ll \! R  \! \ll \! \xi$, 
one has $s \sim R^4$, which is independent of $d$. 

For $\lambda >2$, we first consider $\lambda \! = \! \lambda_p \! = \! d/3$, 
corresponding to the CG-percolation scaling window of width ${\cal O}(L^{-\lambda_p})$, 
and expect that the FK-Ising clusters exhibit nearly the same 
geometric properties as those in the standard bond percolation model on high-d tori. 
The unwrapped correlation length grows faster than $L$ as $\xi_u \! \sim \! L^{\lambda/2} 
\! = \! L^{d/6}$. All the clusters, including $C_1$, scale 
as $s \sim R_u^4$ until the cut-off size $s_\lambda \sim \xi_u^4 \sim L^{2d/3}$.
The Fisher exponent is $\tau=5/2$, and the number of clusters of cut-off sizes is $\scrO(1)$.
In other words, as the standard percolation clusters in high dimensions, 
the FK-Ising clusters manage to keep their shape to be ``thin'' by avoiding touching each other 
and wrapping around the tori for a diverging number of times ($\xi_u/L= L^{d/6-1})$.
This is indicated by Figs.~\ref{fig:C2_DF}, \ref{fig:C12_DF}, \ref{fig:Nspan}(c) 
and other plots in Sec.~\ref{results1}.
It is interesting to note that, in terms of the unwrapped distance $r_u$, 
the two-point correlation function decays as Gaussian-like as $g(r,L) \sim r_u^{2-d}$
until the unwrapped correlation length,
giving $\chi \! \sim \! \xi_u^2 \! \sim \! L^{d/3}$~\cite{Grimm2017Geometric,Grimm2018Finitesize}. 
 The situations for $ \lambda_g \! < \! \lambda \! < \! \lambda_p$ are similar to those for 
the percolation scaling window. 
We have $\xi_u \! \sim \! L^{\lambda/2}$, $s_\lambda \! \sim \! \xi_u^4 \! \sim \! L^{2 \lambda}$ 
and $\chi \! \sim \! \xi_u^2 \! \sim \! L^{\lambda}$. 

\begin{figure}[t]
\centering
\includegraphics[width=0.52\textwidth]{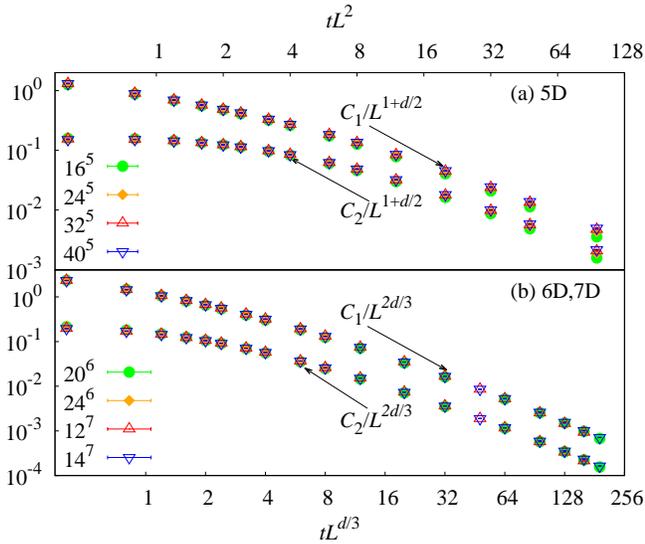}
\caption{The Gaussian and CG-percolation scaling windows 
illustrated by $C_1$ and $C_2$ on high-d tori with (a) $d=5$ and (b) $d=6,7$. 
}
\label{fig:SWC12L13}
\end{figure} 

For $ \lambda \in (d/3, d/2)$, 
we argue that the scaling of the unwrapped diameter of the largest cluster 
becomes $\lambda$-independent as $R_1 \sim L^{d/6}$,
as for $\lambda=\lambda_p$ and $\lambda \geq d/2$. 
Meanwhile, $C_1$ becomes ``fat'' by merging the second-largest and other clusters,
and, from the FSS of $\chi$,
we have $C_1 \sim L^{d_\textsc{l1}} \sim R_1^{d_\textsc{f1}}$, 
with $d_\textsc{l1}=(d+\lambda)/2$ and $d_\textsc{f1}=4+ \delta_\lambda $, 
where $\delta_\lambda \equiv \lambda/\lambda_p -1$ is introduced. 

We expect that all the medium-sized clusters scale as $s \sim R^4$ until the cut-off size $s_\lambda$. 
Since all clusters with $s > s_\lambda$ are merged into the largest cluster, 
$s_\lambda$ should decrease as $\lambda$ increases. 
For the second-largest cluster $C_2$, 
by assuming a linear interpolation between $d_\textsc{l2} =2d/3$ 
for $\lambda=\lambda_p$ and $d_\textsc{l2} =1+d/2$ for $\lambda=d/2$, 
we obtain the finite-size fractal dimension $d_\textsc{l2}=(d-\lambda)+2 \delta_\lambda$. 
With the assumption $C_2 \sim R_2^4$, we have $R_2 \sim L^{d_\textsc{l2}/4}$, 
which is also divergent for $d > 6$.
An argument for $d_\textsc{l2} \geq d-\lambda$ can be provided as following. 
As $\lambda$ increases from $\lambda_p$, all the clusters larger than 
$s_\lambda \! \sim \! C_2 \! \sim \! L^{d_\textsc{l2}}$ are merged into the largest cluster, 
contributing to a total size of $L^d \int_{s_\lambda } s n(s,L) \diff s
\sim L^{d-d_\textsc{l2}/2}$. 
Thus, we have $ L^{d-d_\textsc{l2}/2} \leq C_1 \sim L^{(d+\lambda)/2}$, 
giving $ d_\textsc{l2} \geq d-\lambda$. 
Note that $\delta_\lambda$ is a $d$-independent constant, 
and thus the lower bound $d-\lambda$ becomes sharper and sharper as $d$ increases. 

\begin{figure}[t]
    \centering
    \includegraphics[width=0.52\textwidth]{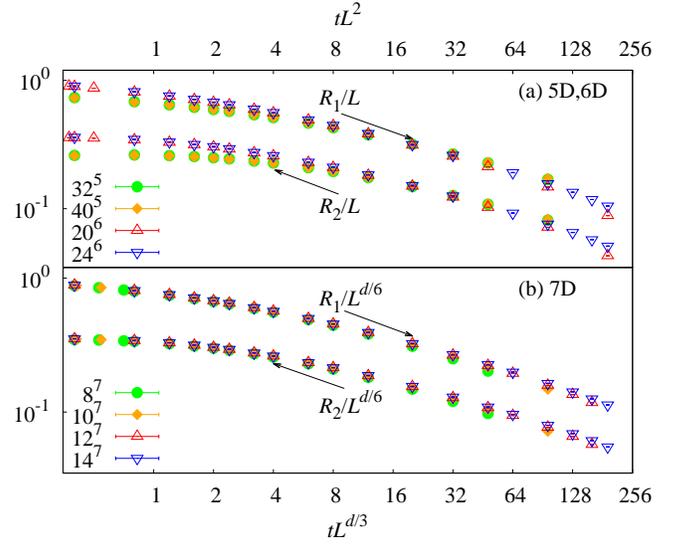}
    \caption{The Gaussian and CG-percolation scaling windows
     illustrated by the unwrapped radii $R_1$ and $R_2$ on high-d tori with (a) $d=5,6$ and (b) $d=7$.}
    \label{fig:SWR12Lh}
\end{figure} 
 
\begin{table}[b]
\centering
\setlength\tabcolsep{1.5pt}
\renewcommand\arraystretch{1.5}
\begin{tabular}{|c|l|ccc|ccc|}
\hline 
\vspace{-1.5mm}
 & & \multicolumn{3}{c|}{largest cluster} & \multicolumn{3}{r|}{second-largest cluster}  
 \\
  $d$     &\;$\lambda \in$  &$d_\textsc{l1}$          &\;\;\;\;$d_\textsc{f1}$\;\;\;\;     & $d_\textsc{r1}$  
                            &$d_\textsc{l2}$          &\;\;\;$d_\textsc{f2}$\;\;\;     & $d_\textsc{r2}$ \\
\hline 
          &$(0,2)$          &$\frac{\lambda}{2}d_g$       &$d_g$              &$\frac{\lambda}{2}$      
                            &$\frac{\lambda}{2}d_g$       &$d_g$              &$\frac{\lambda}{2}$ \\
\cline{2-8} 
$[4,6)$   &$[2,\frac{d}{2})$  &$\frac{\lambda+d}{2}$  &$d_\textsc{l1}$ &1          
                              &$d_g$                  &$d_\textsc{l2}$     &1 \\
\cline{2-8} 
          &$[\frac{d}{2},\infty)$ &$\frac{3}{4}d$                 &\;$d_\textsc{l1}$    &1         
                              &$d_g$                  &$d_\textsc{l2}$    &1 \\
\hline
          &$(0,\frac{d}{3})$  &$2\lambda$             &4                  &$\frac{\lambda}{2}$   
                              &$2\lambda$       &4                  &$\frac{\lambda}{2}$ \\
\cline{2-8} 

$[6,\infty)$ &$[\frac{d}{3},\frac{d}{2})$ &$\frac{\lambda+d}{2}$\;  &$4\!+\!\delta_\lambda$  &$\frac{1}{6}d$  
&\;$d\!-\!\lambda\!+\!2 \delta_\lambda$        &4                  &$\frac{1}{4}d_\textsc{l2}$ \\
\cline{2-8} 
         & $[\frac{d}{2},\infty)\;$ &$\frac{3}{4}d$      &$\frac{9}{2}$       &$\frac{1}{6}d$         
                              &$d_g$                  &4                    &$\frac{1}{4}d_\textsc{l2}$  \\
\hline            
\end{tabular}
\caption{Finite-size and thermodynamic fractal dimensions, $d_{\textsc{l}i}$ and $d_{\textsc{f}i}$,
for the largest and second-largest clusters $(i=1,2)$, 
as the critical point is approached in a speed of $\scrO(L^{-\lambda})$ from 
the high-temperature side. 
The unwrapped diameters of clusters scale as $R_i \sim L^{d_{\textsc{r}i}}$,
with $d_{\textsc{r}i}=d_{\textsc{l}i}/d_{\textsc{f}i}$
For clarity, we use $d_g=d/2+1$ 
and $\delta_\lambda = \lambda/\lambda_p-1$ with $\lambda_p=d/3$. 
Note that the exact values of many of these exponents are conjectured 
on the basis of numerics, insights from RG theory and CG asymptotics, 
or even from linear interpolation. 
}
\label{tab:SWdall_finite}
\end{table}

In a brief summary, as the critical point is approached as $\scrO(L^{-\lambda})$ from 
 the high-T side, the FK-Ising model exhibits the simultaneous existence of 
 the CG-Ising scaling window of width $\scrO(L^{-d/2})$ and of the Gaussian scaling window
 of width $\scrO(L^{-2})$. For $d \geq 6$, in between, there exists another scaling window of width $\scrO(L^{-d/3})$, corresponding to the CG-percolation scaling window. 
 Since the high-d percolation exhibits both the Gaussian and CG-percolation scaling windows, 
 we say that the FK-Ising model for $d \geq 6$ exhibits the simultaneous existence 
 of the high-d percolation and the CG-Ising scaling windows. 
 As $\lambda$ increases, the correlation length $\xi$ reaches the order of $L$ already in 
 the Gaussian scaling window.
 For $ 4 \! \leq \! d \! < \! 6$,  $\xi \sim L$ saturates as long as $\lambda >2$, and the largest and second-largest clusters display different geometric properties. 
 For $d > 6$, the correlation length in an unwrapped way saturates 
 in the high-d percolation scaling window as $\xi_u \sim L^{d/6}$, 
 and the two-length-scale behaviors develop for $\lambda > d/3$. 
 As an illustration, Figs.~\ref{fig:SWC12L12}, \ref{fig:SWC12L13} and~\ref{fig:SWR12Lh}
 display the FSS behaviors of the largest and second-largest clusters 
 in the CG-Ising, Gaussian, and CG-percolation scaling windows. 
 Table~\ref{tab:SWdall_finite} lists the exact values of critical exponents 
 for different $\lambda$ and $d$, 
 including the finite-size and thermodynamic fractal dimensions 
 as well as the scaling exponents for the unwrapped cluster diameters.

\subsection{Low-temperature side}
\label{low-T asymptotics} 
We then consider that the critical point is approached from the low-T side, i.e., $a<0$. 
When the temperature is decreased, more and more clusters merge into the largest cluster, such that the second-largest cluster $C_2$ becomes smaller and smaller.
One would not expect to observe the percolation scaling windows with $\lambda=d/3$, in which $C_2 \sim L^{2d/3}$.
Recall that it was observed that the Ising model on the CG has a critical window with a width of order ${\cal O}(V^{-1/2})$ both for the spin representation and FK representation~\cite{Luijten1997Interaction,Luczak2006Phase}. Thus, one would expect that within this critical window, the FSS behaviors are the same  as  those at criticality. In other words, for $\lambda \in [d/2,\infty)$, the scaling behavior is the same for the high-T approach and low-T approach, and the corresponding exponents are listed in Table~\ref{tab:SWdall_finite}. In Fig.~\ref{fig:SWC12L12}, we plot the rescaled cluster sizes $C_1/L^{3d/4}$ and $C_2 \ln L/L^{1+d/2}$  versus $tL^{d/2}$. The good data collapse gives solid support to the existence of the critical window.

\subsection{Crossover to the thermodynamic limit}
\label{Asymmetric behaviors} 
\begin{figure}[t]
	\centering
	\includegraphics[width=0.48\textwidth]{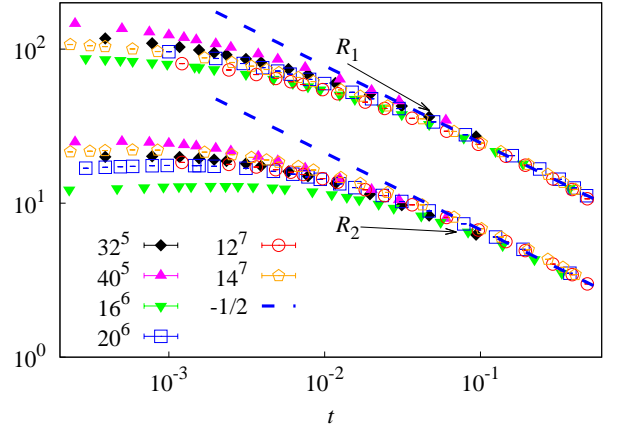}
	\caption{The thermodynamic scaling behavior of the radius $R_1$ and $R_2$ from the high-T approach for $d=5,6,7$.}
	\label{fig:SWR12th}
\end{figure} 
From above, we find there is an asymmetric FSS behavior from the high-T  and low-T approaches, which is unconventional in most critical systems. Here, we note that it  not only affects the FSS behavior, but also the thermodynamic scaling behavior.  \par  

Like the susceptibility $\chi$ in Eq.~\eqref{eq:chi_FSS}, we assume that the FSS scaling behaviors of other quantities obey a similar form to   
\begin{equation}
              {\cal Q}(t, L) = L^{y_{\scrQ} \cdot  \lambda_t} \tilde{\scrQ}(tL^{\lambda_t}).
        \end{equation}
Moreover, under the condition that the quantity ${\cal Q}$ is well defined directly in the thermodynamic limit, we assume that the scaling function follows 
        \begin{equation}
 \tilde{\cal Q}(x) \sim 
    \begin{cases}
         {\rm{const}}, \; x \to 0, \\ 
         |x|^{-y_{\cal Q}},  x \to \infty, \\ 
    \end{cases}
\end{equation}
such that ${\cal Q}(t)$ returns to the thermodynamic scaling behavior ${\cal Q}(t) \sim |t|^{-y_{\scrQ}}$. For $\chi$, the exponents $y_{\scrQ}=\gamma=1$ and $\lambda_t=d/2$. 
We then take the radii $R_1$ and $R_2$ as an example and  consider the correlation-length exponents $\nu_1$ and $\nu_2$, which are defined as $R_1(t) \sim |t|^{-\nu_1}$ and $R_2(t) \sim |t|^{-\nu_2}$. 
We note that $\nu_2$ can be well defined both from the high-T and low-T sides, while $\nu_1$ is only well defined from the high-T side. \par 

We first consider the high-T approach. We assume that the radii $R_1$ and $R_2$ have the same scaling behavior $R_{1,2}(t,L) \sim L \tilde{R}_{1,2}(tL^{2})$ for $4<d<6$, such that we obtain $\nu_1 = \nu_2 =1/2$. In this case, it is consistent  with the scaling behavior in Table~\ref{tab:SWdall_finite}. In other words, if one takes $t\sim L^{-\lambda}$ it turns to be $R_{1,2} \sim L^{\lambda/2}$ for $\lambda \leq \lambda_t =2$ and $R_{1,2} \sim L$ for $\lambda > \lambda_t$. 
For $d \ge 6$, following a similar procedure, we assume the scaling behaviors $R_{1,2} \sim L^{d/6} \tilde{R}_{1,2}(tL^{d/3})$ and $R_{1,2} (t) \sim |t|^{-1/2}$. 
Figure~\ref{fig:SWR12th} illustrates the above thermodynamic scaling behaviors of $R_1$ and $R_2$. \par 

We then consider that it approaches the critical point from the low-T side. Since one has $R_2 \sim L \tilde{R}_2(tL^{d/2})$ for $4<d<6$ and $R_2\sim L^{1/4+d/8} \tilde{R}_2(tL^{d/2})$ for $d \ge 6$, it is expected that $R_2(t) \sim |t|^{-\nu_2}$ with $\nu_2 =2/d$  for $4<d<6$ and $\nu_2=1/4+1/2d$ for $d\ge 6$. As $d\to \infty$, the exponent $\nu_2$  reduces to $1/4$, consistent with the calculation on the Bethe lattice~\cite{Chayes1999Meanfield}.

\begin{figure}[t]
	\centering
	\includegraphics[width=0.50\textwidth]{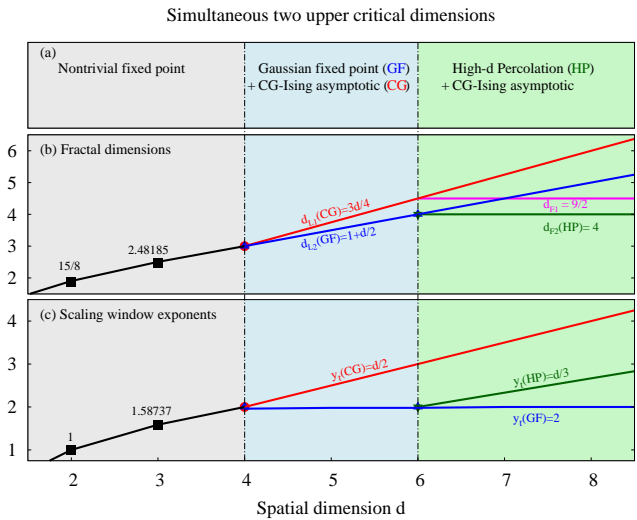}
	\caption{Demonstration of the simultaneous two upper critical dimensions of the Ising model in its FK representation. The scaling behaviors are governed by nontrivial fixed points for $d<4$, by the Gaussian fixed point asymptotics and complete graph asymptotics for $4<d<6$, and by the  complete graph asymptotics and high-d percolation asymptotic for $d \ge 6$, as illustrated in (a). The values of the finite-size fractal dimension $d_\textsc{l}$ and of the thermodynamic fractal dimension $d_\textsc{f}$ are given in (b) for the largest cluster and second-largest cluster. The value of the thermal-like renormalization exponent $y_t$, governing the size of the corresponding scaling window as ${\cal O}(L^{-y_t})$, is given in (c).  }
	\label{fig:DFL}
\end{figure}

\section{Discussion}
\label{discussion} 

In this work, we perform large-scale Monte Carlo simulations on high-dimensional (high-d) tori and the complete graph (CG). Based on our numerical results, we  provide a detailed and complete report to support our conjecture of the simultaneous existence of the two upper critical dimensions ($d_c=4, d_p=6$) in the Fortuin-Kasteleyn (FK) representation of the Ising model. 
Other rich phenomena are further observed. In particular, as long as $d > d_c$, there are two configuration sectors, two-length-scale behaviors, and two scaling windows. The scaling behaviors are conjectured to be governed by the Gaussian fixed point (GFP) asymptotics and CG-Ising asymptotics for $4<d<6$, and by the CG-Ising asymptotics and high-d percolation asymptotics for $d\ge 6$.
For clarity, the $d$-dependent values of various critical exponents are summarized in Fig.~\ref{fig:DFL}. 
It is unexpected at first glance that for $d \ge d_p$, many scaling behaviors of the FK-Ising clusters are the same  as the critical high-d percolation, including the thermodynamic fractal dimension $d_\textsc{f2} =4$ and  the scaling behavior of the radius $R_1 \sim L^{d/6}$ and the number of spanning clusters $N_s \sim L^{d-6}$.  \par

The rich phenomena observed in the FK representation deepen our understanding of the Ising model.  
A natural question is whether these  phenomena can be 
 observed in other representations. For the spin representation, critical behaviors are simpler:
no percolation-like behaviors exist and the upper critical dimension $d_p = 6$ does not exist. 
Apart from the FK representation, there is another geometric  representation, i.e., the loop representation, which can be linked to the FK representation in the framework of the loop-cluster joint model~\cite{Zhang2020Loop}. 
Recently, it was shown in Ref.~\cite{Li2023Geometric} that studying the loop representation on the CG within the framework of the loop-cluster joint model provides a natural and simple explanation for the appearance  of percolation behaviors in the FK Ising model, which results in the existence of two length scales, two configuration sectors, and two scaling windows. 
Thus, we would expect that the study of  the loop representation on high-dimensional tori, which is still under our investigation, would provide some explanations for the observations in the FK Ising model on tori, especially for the two upper critical dimensions.  \par 

Here, we emphasize that the general scenario in Fig.~\ref{fig:DFL} is just a conjecture based on extensive simulations and insights from exact CG solutions and the existing RG calculations.
Further studies are needed to judge the validity of the conjectured scenario in Fig.~\ref{fig:DFL} as well as the values of the critical exponents.


\section*{Acknowledgements}
	This work has been supported by the National Natural Science Foundation of China 
(under Grant No. 12275263), the National Key R\&D Program of China (under Grant No. 2018YFA0306501).
We thank Eren M. El\c{c}i, Jens Grimm, Timothy Garoni, Martin Weigel, Jonathan Machta, and Jesper Jacobsen  for valuable discussions.


\begin{thebibliography}{50}%
	\makeatletter
	\providecommand \@ifxundefined [1]{%
		\@ifx{#1\undefined}
	}%
	\providecommand \@ifnum [1]{%
		\ifnum #1\expandafter \@firstoftwo
		\else \expandafter \@secondoftwo
		\fi
	}%
	\providecommand \@ifx [1]{%
		\ifx #1\expandafter \@firstoftwo
		\else \expandafter \@secondoftwo
		\fi
	}%
	\providecommand \natexlab [1]{#1}%
	\providecommand \enquote  [1]{``#1''}%
	\providecommand \bibnamefont  [1]{#1}%
	\providecommand \bibfnamefont [1]{#1}%
	\providecommand \citenamefont [1]{#1}%
	\providecommand \href@noop [0]{\@secondoftwo}%
	\providecommand \href [0]{\begingroup \@sanitize@url \@href}%
	\providecommand \@href[1]{\@@startlink{#1}\@@href}%
	\providecommand \@@href[1]{\endgroup#1\@@endlink}%
	\providecommand \@sanitize@url [0]{\catcode `\\12\catcode `\$12\catcode
		`\&12\catcode `\#12\catcode `\^12\catcode `\_12\catcode `\%12\relax}%
	\providecommand \@@startlink[1]{}%
	\providecommand \@@endlink[0]{}%
	\providecommand \url  [0]{\begingroup\@sanitize@url \@url }%
	\providecommand \@url [1]{\endgroup\@href {#1}{\urlprefix }}%
	\providecommand \urlprefix  [0]{URL }%
	\providecommand \Eprint [0]{\href }%
	\providecommand \doibase [0]{https://doi.org/}%
	\providecommand \selectlanguage [0]{\@gobble}%
	\providecommand \bibinfo  [0]{\@secondoftwo}%
	\providecommand \bibfield  [0]{\@secondoftwo}%
	\providecommand \translation [1]{[#1]}%
	\providecommand \BibitemOpen [0]{}%
	\providecommand \bibitemStop [0]{}%
	\providecommand \bibitemNoStop [0]{.\EOS\space}%
	\providecommand \EOS [0]{\spacefactor3000\relax}%
	\providecommand \BibitemShut  [1]{\csname bibitem#1\endcsname}%
	\let\auto@bib@innerbib\@empty
	\bibitem [{\citenamefont {Friedli}\ and\ \citenamefont
		{Velenik}(2017)}]{friedli2017statistical}%
	\BibitemOpen
	\bibfield  {author} {\bibinfo {author} {\bibfnamefont {S.}~\bibnamefont
			{Friedli}}\ and\ \bibinfo {author} {\bibfnamefont {Y.}~\bibnamefont
			{Velenik}},\ }\href@noop {} {\emph {\bibinfo {title} {Statistical Mechanics
				of Lattice Systems: A Concrete Mathematical Introduction}}}\ (\bibinfo
	{publisher} {Cambridge University Press},\ \bibinfo {year}
	{2017})\BibitemShut {NoStop}%
	\bibitem [{\citenamefont {Lenz}(1920)}]{lenz1920beitrag}%
	\BibitemOpen
	\bibfield  {author} {\bibinfo {author} {\bibfnamefont {W.}~\bibnamefont
			{Lenz}},\ }\bibfield  {title} {\bibinfo {title} {Beitrag zum verst{\"a}ndnis
			der magnetischen erscheinungen in festen k{\"o}rpern},\ }\href
	{}
	{\bibfield  {journal} {\bibinfo  {journal} {European Physical Journal A}\
		}\textbf {\bibinfo {volume} {21}},\ \bibinfo {pages} {613} (\bibinfo {year}
		{1920})}\BibitemShut {NoStop}%
	\bibitem [{\citenamefont {Ising}(1925)}]{Ising1925Beitrag}%
	\BibitemOpen
	\bibfield  {author} {\bibinfo {author} {\bibfnamefont {E.}~\bibnamefont
			{Ising}},\ }\bibfield  {title} {\bibinfo {title} {{Beitrag zur Theorie des
				Ferromagnetismus}},\ }\href
	{} {\bibfield  {journal}
		{\bibinfo  {journal} {Zeitschrift f\"ur Physik}\ }\textbf {\bibinfo {volume}
			{31}},\ \bibinfo {pages} {253} (\bibinfo {year} {1925})}\BibitemShut
	{NoStop}%
	\bibitem [{\citenamefont {Onsager}(1944)}]{Onsager1944Crystal}%
	\BibitemOpen
	\bibfield  {author} {\bibinfo {author} {\bibfnamefont {L.}~\bibnamefont
			{Onsager}},\ }\bibfield  {title} {\bibinfo {title} {Crystal statistics. {I.
				A} two-dimensional model with an order-disorder transition},\ }\href
	{} {\bibfield  {journal}
		{\bibinfo  {journal} {Physical Review}\ }\textbf {\bibinfo {volume} {65}},\
		\bibinfo {pages} {117} (\bibinfo {year} {1944})}\BibitemShut {NoStop}%
	\bibitem [{\citenamefont {Kramers}\ and\ \citenamefont
		{Wannier}(1941)}]{Kramers1941statistics}%
	\BibitemOpen
	\bibfield  {author} {\bibinfo {author} {\bibfnamefont {H.~A.}\ \bibnamefont
			{Kramers}}\ and\ \bibinfo {author} {\bibfnamefont {G.~H.}\ \bibnamefont
			{Wannier}},\ }\bibfield  {title} {\bibinfo {title} {Statistics of the
			two-dimensional ferromagnet. {Part I}},\ }\href
	{} {\bibfield  {journal}
		{\bibinfo  {journal} {Physical Review}\ }\textbf {\bibinfo {volume} {60}},\
		\bibinfo {pages} {252} (\bibinfo {year} {1941})}\BibitemShut {NoStop}%
	\bibitem [{\citenamefont {Baxter}(2007)}]{Baxter2007}%
	\BibitemOpen
	\bibfield  {author} {\bibinfo {author} {\bibfnamefont {R.}~\bibnamefont
			{Baxter}},\ }\href@noop {} {\emph {\bibinfo {title} {Exactly Solved Models in
				Statistical Mechanics}}},\ Dover books on physics\ (\bibinfo  {publisher}
	{Dover, New York},\ \bibinfo {year} {2007})\BibitemShut {NoStop}%
	\bibitem [{\citenamefont {Yang}(1952)}]{Yang1952Spontaneous}%
	\BibitemOpen
	\bibfield  {author} {\bibinfo {author} {\bibfnamefont {C.~N.}\ \bibnamefont
			{Yang}},\ }\bibfield  {title} {\bibinfo {title} {The spontaneous
			magnetization of a two-dimensional {Ising} model},\ }\href
	{} {\bibfield  {journal}
		{\bibinfo  {journal} {Physical Review}\ }\textbf {\bibinfo {volume} {85}},\
		\bibinfo {pages} {808} (\bibinfo {year} {1952})}\BibitemShut {NoStop}%
	\bibitem [{\citenamefont {Aizenman}\ and\ \citenamefont
		{Fern\'andez}(1986)}]{AizenmanFernandez1986}%
	\BibitemOpen
	\bibfield  {author} {\bibinfo {author} {\bibfnamefont {M.}~\bibnamefont
			{Aizenman}}\ and\ \bibinfo {author} {\bibfnamefont {R.}~\bibnamefont
			{Fern\'andez}},\ }\bibfield  {title} {\bibinfo {title} {On the critical
			behavior of the magnetization in high-dimensional {Ising} models},\ }\href
	{} {\bibfield  {journal}
		{\bibinfo  {journal} {Journal of Statistical Physics}\ }\textbf {\bibinfo
			{volume} {44}},\ \bibinfo {pages} {393–454} (\bibinfo {year}
		{1986})}\BibitemShut {NoStop}%
	\bibitem [{\citenamefont {Aizenman}\ \emph {et~al.}(2015)\citenamefont
		{Aizenman}, \citenamefont {Duminil-Copin},\ and\ \citenamefont
		{Sidoravicius}}]{AizenmanDuminilCopinVladas2015}%
	\BibitemOpen
	\bibfield  {author} {\bibinfo {author} {\bibfnamefont {M.}~\bibnamefont
			{Aizenman}}, \bibinfo {author} {\bibfnamefont {H.}~\bibnamefont
			{Duminil-Copin}},\ and\ \bibinfo {author} {\bibfnamefont {V.}~\bibnamefont
			{Sidoravicius}},\ }\bibfield  {title} {\bibinfo {title} {Random currents and
			continuity of {Ising} model’s spontaneous magnetization},\ }\href
	{} {\bibfield
		{journal} {\bibinfo  {journal} {Communications in Mathematical Physics}\
		}\textbf {\bibinfo {volume} {334}},\ \bibinfo {pages} {719–742} (\bibinfo
		{year} {2015})}\BibitemShut {NoStop}%
	\bibitem [{\citenamefont {Deng}\ and\ \citenamefont
		{Bl{\"o}te}(2003)}]{Deng2003Simultaneous}%
	\BibitemOpen
	\bibfield  {author} {\bibinfo {author} {\bibfnamefont {Y.}~\bibnamefont
			{Deng}}\ and\ \bibinfo {author} {\bibfnamefont {H.~W.~J.}\ \bibnamefont
			{Bl{\"o}te}},\ }\bibfield  {title} {\bibinfo {title} {Simultaneous analysis
			of several models in the three-dimensional {Ising} universality class},\
	}\href {} {\bibfield
		{journal} {\bibinfo  {journal} {Physical Review E}\ }\textbf {\bibinfo
			{volume} {68}},\ \bibinfo {pages} {036125} (\bibinfo {year}
		{2003})}\BibitemShut {NoStop}%
	\bibitem [{\citenamefont {Ferrenberg}\ \emph {et~al.}(2018)\citenamefont
		{Ferrenberg}, \citenamefont {Xu},\ and\ \citenamefont
		{Landau}}]{Ferrenberg2018Pushing}%
	\BibitemOpen
	\bibfield  {author} {\bibinfo {author} {\bibfnamefont {A.~M.}\ \bibnamefont
			{Ferrenberg}}, \bibinfo {author} {\bibfnamefont {J.}~\bibnamefont {Xu}},\
		and\ \bibinfo {author} {\bibfnamefont {D.~P.}\ \bibnamefont {Landau}},\
	}\bibfield  {title} {\bibinfo {title} {Pushing the limits of {{Monte Carlo}}
			simulations for the three-dimensional {{Ising}} model},\ }\href
	{} {\bibfield
		{journal} {\bibinfo  {journal} {Physical Review E}\ }\textbf {\bibinfo
			{volume} {97}},\ \bibinfo {pages} {043301} (\bibinfo {year}
		{2018})}\BibitemShut {NoStop}%
	\bibitem [{\citenamefont {Hou}\ \emph {et~al.}(2019)\citenamefont {Hou},
		\citenamefont {Fang}, \citenamefont {Wang}, \citenamefont {Hu},\ and\
		\citenamefont {Deng}}]{Hou2019Geometric}%
	\BibitemOpen
	\bibfield  {author} {\bibinfo {author} {\bibfnamefont {P.}~\bibnamefont
			{Hou}}, \bibinfo {author} {\bibfnamefont {S.}~\bibnamefont {Fang}}, \bibinfo
		{author} {\bibfnamefont {J.}~\bibnamefont {Wang}}, \bibinfo {author}
		{\bibfnamefont {H.}~\bibnamefont {Hu}},\ and\ \bibinfo {author}
		{\bibfnamefont {Y.}~\bibnamefont {Deng}},\ }\bibfield  {title} {\bibinfo
		{title} {Geometric properties of the {Fortuin-Kasteleyn} representation of
			the {Ising} model},\ }\href
	{} {\bibfield  {journal}
		{\bibinfo  {journal} {Physical Review E}\ }\textbf {\bibinfo {volume} {99}},\
		\bibinfo {pages} {042150} (\bibinfo {year} {2019})}\BibitemShut {NoStop}%
	\bibitem [{Gui()}]{Guida19973dIsing}%
	\BibitemOpen
	\href@noop {} {\ }\BibitemShut {NoStop}%
	\bibitem [{\citenamefont {Kos}\ \emph {et~al.}(2016)\citenamefont {Kos},
		\citenamefont {Poland}, \citenamefont {{Simmons-Duffin}},\ and\ \citenamefont
		{Vichi}}]{Kos2016Precision}%
	\BibitemOpen
	\bibfield  {author} {\bibinfo {author} {\bibfnamefont {F.}~\bibnamefont
			{Kos}}, \bibinfo {author} {\bibfnamefont {D.}~\bibnamefont {Poland}},
		\bibinfo {author} {\bibfnamefont {D.}~\bibnamefont {{Simmons-Duffin}}},\ and\
		\bibinfo {author} {\bibfnamefont {A.}~\bibnamefont {Vichi}},\ }\bibfield
	{title} {\bibinfo {title} {Precision islands in the {Ising} and {O(N)}
			models},\ }\href {}
	{\bibfield  {journal} {\bibinfo  {journal} {Journal of High Energy Physics}\
		}\textbf {\bibinfo {volume} {2016}},\ \bibinfo {pages} {36} (\bibinfo {year}
		{2016})}\BibitemShut {NoStop}%
	\bibitem [{\citenamefont {Fernandez}\ \emph {et~al.}(2013)\citenamefont
		{Fernandez}, \citenamefont {Fr{\"o}hlich},\ and\ \citenamefont
		{Sokal}}]{FernandezFrohlichSokal13}%
	\BibitemOpen
	\bibfield  {author} {\bibinfo {author} {\bibfnamefont {R.}~\bibnamefont
			{Fernandez}}, \bibinfo {author} {\bibfnamefont {J.}~\bibnamefont
			{Fr{\"o}hlich}},\ and\ \bibinfo {author} {\bibfnamefont {A.}~\bibnamefont
			{Sokal}},\ }\href@noop {} {\emph {\bibinfo {title} {Random walks, critical
				phenomena, and triviality in quantum field theory}}},\ Theoretical and
	Mathematical Physics\ (\bibinfo  {publisher} {Springer Berlin Heidelberg},\
	\bibinfo {year} {2013})\BibitemShut {NoStop}%
	\bibitem [{\citenamefont {Grimmett}(2006)}]{Grimmett2006Random}%
	\BibitemOpen
	\bibfield  {author} {\bibinfo {author} {\bibfnamefont {G.~R.}\ \bibnamefont
			{Grimmett}},\ }\href@noop {} {\emph {\bibinfo {title} {The random-cluster
				model}}},\ Vol.\ \bibinfo {volume} {333}\ (\bibinfo  {publisher} {Springer
		Science \& Business Media, Berlin},\ \bibinfo {year} {2006})\BibitemShut
	{NoStop}%
	\bibitem [{\citenamefont {Wu}(1982)}]{Wu1982The}%
	\BibitemOpen
	\bibfield  {author} {\bibinfo {author} {\bibfnamefont {F.~Y.}\ \bibnamefont
			{Wu}},\ }\bibfield  {title} {\bibinfo {title} {The {Potts} model},\ }\href
	{} {\bibfield  {journal}
		{\bibinfo  {journal} {Reviews of Modern Physics}\ }\textbf {\bibinfo {volume}
			{54}},\ \bibinfo {pages} {235} (\bibinfo {year} {1982})}\BibitemShut
	{NoStop}%
	\bibitem [{\citenamefont {Stauffer}\ and\ \citenamefont
		{Aharony}(2018)}]{Stauffer2018Introduction}%
	\BibitemOpen
	\bibfield  {author} {\bibinfo {author} {\bibfnamefont {D.}~\bibnamefont
			{Stauffer}}\ and\ \bibinfo {author} {\bibfnamefont {A.}~\bibnamefont
			{Aharony}},\ }\href@noop {} {\emph {\bibinfo {title} {Introduction to
				percolation theory}}}\ (\bibinfo  {publisher} {CRC press},\ \bibinfo {year}
	{2018})\BibitemShut {NoStop}%
	\bibitem [{\citenamefont {Wolff}(1989)}]{Wolff1989Collective}%
	\BibitemOpen
	\bibfield  {author} {\bibinfo {author} {\bibfnamefont {U.}~\bibnamefont
			{Wolff}},\ }\bibfield  {title} {\bibinfo {title} {Collective {Monte Carlo}
			updating for spin systems},\ }\href
	{} {\bibfield  {journal}
		{\bibinfo  {journal} {Physical Review Letters}\ }\textbf {\bibinfo {volume}
			{62}},\ \bibinfo {pages} {361} (\bibinfo {year} {1989})}\BibitemShut
	{NoStop}%
	\bibitem [{\citenamefont {Swendsen}\ and\ \citenamefont
		{Wang}(1987)}]{Swendsen1987Nonuniversal}%
	\BibitemOpen
	\bibfield  {author} {\bibinfo {author} {\bibfnamefont {R.~H.}\ \bibnamefont
			{Swendsen}}\ and\ \bibinfo {author} {\bibfnamefont {J.-S.}\ \bibnamefont
			{Wang}},\ }\bibfield  {title} {\bibinfo {title} {Nonuniversal critical
			dynamics in {Monte Carlo} simulations},\ }\href
	{} {\bibfield  {journal}
		{\bibinfo  {journal} {Physical Review Letters}\ }\textbf {\bibinfo {volume}
			{58}},\ \bibinfo {pages} {86} (\bibinfo {year} {1987})}\BibitemShut {NoStop}%
	\bibitem [{\citenamefont {Zhang}\ \emph {et~al.}(2020)\citenamefont {Zhang},
		\citenamefont {Michel}, \citenamefont {El{\c c}i},\ and\ \citenamefont
		{Deng}}]{Zhang2020Loop}%
	\BibitemOpen
	\bibfield  {author} {\bibinfo {author} {\bibfnamefont {L.}~\bibnamefont
			{Zhang}}, \bibinfo {author} {\bibfnamefont {M.}~\bibnamefont {Michel}},
		\bibinfo {author} {\bibfnamefont {E.~M.}\ \bibnamefont {El{\c c}i}},\ and\
		\bibinfo {author} {\bibfnamefont {Y.}~\bibnamefont {Deng}},\ }\bibfield
	{title} {\bibinfo {title} {Loop-cluster coupling and algorithm for classical
			statistical models},\ }\href
	{} {\bibfield
		{journal} {\bibinfo  {journal} {Physical Review Letters}\ }\textbf {\bibinfo
			{volume} {125}},\ \bibinfo {pages} {200603} (\bibinfo {year}
		{2020})}\BibitemShut {NoStop}%
	\bibitem [{\citenamefont {Zia}\ and\ \citenamefont
		{Wallace}(1975)}]{Zia1975Critical}%
	\BibitemOpen
	\bibfield  {author} {\bibinfo {author} {\bibfnamefont {R.~K.~P.}\
			\bibnamefont {Zia}}\ and\ \bibinfo {author} {\bibfnamefont {D.~J.}\
			\bibnamefont {Wallace}},\ }\bibfield  {title} {\bibinfo {title} {Critical
			behaviour of the continuous n-component {Potts} model},\ }\href
	{} {\bibfield
		{journal} {\bibinfo  {journal} {Journal of Physics A: Mathematical and
				General}\ }\textbf {\bibinfo {volume} {8}},\ \bibinfo {pages} {1495}
		(\bibinfo {year} {1975})}\BibitemShut {NoStop}%
	\bibitem [{\citenamefont {Amit}(1976)}]{Amit1976Renormalization}%
	\BibitemOpen
	\bibfield  {author} {\bibinfo {author} {\bibfnamefont {D.~J.}\ \bibnamefont
			{Amit}},\ }\bibfield  {title} {\bibinfo {title} {Renormalization of the
			{Potts} model},\ }\href
	{} {\bibfield
		{journal} {\bibinfo  {journal} {Journal of Physics A: Mathematical and
				General}\ }\textbf {\bibinfo {volume} {9}},\ \bibinfo {pages} {1441}
		(\bibinfo {year} {1976})}\BibitemShut {NoStop}%
	\bibitem [{\citenamefont {Chayes}\ and\ \citenamefont
		{Chayes}(1987)}]{Chayes1987Upper}%
	\BibitemOpen
	\bibfield  {author} {\bibinfo {author} {\bibfnamefont {J.~T.}\ \bibnamefont
			{Chayes}}\ and\ \bibinfo {author} {\bibfnamefont {L.}~\bibnamefont
			{Chayes}},\ }\bibfield  {title} {\bibinfo {title} {On the upper critical
			dimension of bernoulli percolation},\ }\href
	{} {\bibfield  {journal} {\bibinfo
			{journal} {Communications in Mathematical Physics}\ }\textbf {\bibinfo
			{volume} {113}},\ \bibinfo {pages} {27} (\bibinfo {year} {1987})}\BibitemShut
	{NoStop}%
	\bibitem [{\citenamefont {Chayes}\ \emph {et~al.}(1999)\citenamefont {Chayes},
		\citenamefont {Coniglio}, \citenamefont {Machta},\ and\ \citenamefont
		{Shtengel}}]{Chayes1999Meanfield}%
	\BibitemOpen
	\bibfield  {author} {\bibinfo {author} {\bibfnamefont {L.}~\bibnamefont
			{Chayes}}, \bibinfo {author} {\bibfnamefont {A.}~\bibnamefont {Coniglio}},
		\bibinfo {author} {\bibfnamefont {J.}~\bibnamefont {Machta}},\ and\ \bibinfo
		{author} {\bibfnamefont {K.}~\bibnamefont {Shtengel}},\ }\bibfield  {title}
	{\bibinfo {title} {Mean-field theory for percolation models of the {Ising}
			type},\ }\href {}
	{\bibfield  {journal} {\bibinfo  {journal} {Journal of Statistical Physics}\
		}\textbf {\bibinfo {volume} {94}},\ \bibinfo {pages} {53} (\bibinfo {year}
		{1999})}\BibitemShut {NoStop}%
	\bibitem [{\citenamefont {Fang}\ \emph {et~al.}(2022)\citenamefont {Fang},
		\citenamefont {Zhou},\ and\ \citenamefont {Deng}}]{Fang2022Geometric}%
	\BibitemOpen
	\bibfield  {author} {\bibinfo {author} {\bibfnamefont {S.}~\bibnamefont
			{Fang}}, \bibinfo {author} {\bibfnamefont {Z.}~\bibnamefont {Zhou}},\ and\
		\bibinfo {author} {\bibfnamefont {Y.}~\bibnamefont {Deng}},\ }\bibfield
	{title} {\bibinfo {title} {Geometric upper critical dimensions of the {Ising}
			model},\ }\href
	{} {\bibfield
		{journal} {\bibinfo  {journal} {Chinese Physics Letters}\ }\textbf {\bibinfo
			{volume} {39}},\ \bibinfo {pages} {080502} (\bibinfo {year}
		{2022})}\BibitemShut {NoStop}%
	\bibitem [{\citenamefont {Br{\'e}zin}\ and\ \citenamefont
		{{Zinn-Justin}}(1985)}]{Brezin1985Finite}%
	\BibitemOpen
	\bibfield  {author} {\bibinfo {author} {\bibfnamefont {E.}~\bibnamefont
			{Br{\'e}zin}}\ and\ \bibinfo {author} {\bibfnamefont {J.}~\bibnamefont
			{{Zinn-Justin}}},\ }\bibfield  {title} {\bibinfo {title} {Finite size effects
			in phase transitions},\ }\href
	{}
	{\bibfield  {journal} {\bibinfo  {journal} {Nuclear Physics B}\ }\textbf
		{\bibinfo {volume} {257}},\ \bibinfo {pages} {867} (\bibinfo {year}
		{1985})}\BibitemShut {NoStop}%
	\bibitem [{\citenamefont {Binder}\ \emph {et~al.}(1985)\citenamefont {Binder},
		\citenamefont {Nauenberg}, \citenamefont {Privman},\ and\ \citenamefont
		{Young}}]{Binder1985Finitesize}%
	\BibitemOpen
	\bibfield  {author} {\bibinfo {author} {\bibfnamefont {K.}~\bibnamefont
			{Binder}}, \bibinfo {author} {\bibfnamefont {M.}~\bibnamefont {Nauenberg}},
		\bibinfo {author} {\bibfnamefont {V.}~\bibnamefont {Privman}},\ and\ \bibinfo
		{author} {\bibfnamefont {A.~P.}\ \bibnamefont {Young}},\ }\bibfield  {title}
	{\bibinfo {title} {Finite-size tests of hyperscaling},\ }\href
	{} {\bibfield  {journal}
		{\bibinfo  {journal} {Physical Review B}\ }\textbf {\bibinfo {volume} {31}},\
		\bibinfo {pages} {1498} (\bibinfo {year} {1985})}\BibitemShut {NoStop}%
	\bibitem [{\citenamefont {Wittmann}\ and\ \citenamefont
		{Young}(2014)}]{Wittmann2014Finitesize}%
	\BibitemOpen
	\bibfield  {author} {\bibinfo {author} {\bibfnamefont {M.}~\bibnamefont
			{Wittmann}}\ and\ \bibinfo {author} {\bibfnamefont {A.~P.}\ \bibnamefont
			{Young}},\ }\bibfield  {title} {\bibinfo {title} {Finite-size scaling above
			the upper critical dimension},\ }\href
	{} {\bibfield  {journal}
		{\bibinfo  {journal} {Physical Review E}\ }\textbf {\bibinfo {volume} {90}},\
		\bibinfo {pages} {062137} (\bibinfo {year} {2014})}\BibitemShut {NoStop}%
	\bibitem [{\citenamefont {{Flores-Sola}}\ \emph {et~al.}(2016)\citenamefont
		{{Flores-Sola}}, \citenamefont {Berche}, \citenamefont {Kenna},\ and\
		\citenamefont {Weigel}}]{Flores-Sola2016Role}%
	\BibitemOpen
	\bibfield  {author} {\bibinfo {author} {\bibfnamefont {E.}~\bibnamefont
			{{Flores-Sola}}}, \bibinfo {author} {\bibfnamefont {B.}~\bibnamefont
			{Berche}}, \bibinfo {author} {\bibfnamefont {R.}~\bibnamefont {Kenna}},\ and\
		\bibinfo {author} {\bibfnamefont {M.}~\bibnamefont {Weigel}},\ }\bibfield
	{title} {\bibinfo {title} {Role of {Fourier} modes in finite-size scaling
			above the upper critical dimension},\ }\href
	{} {\bibfield
		{journal} {\bibinfo  {journal} {Physical Review Letters}\ }\textbf {\bibinfo
			{volume} {116}},\ \bibinfo {pages} {115701} (\bibinfo {year}
		{2016})}\BibitemShut {NoStop}%
	\bibitem [{\citenamefont {Grimm}\ \emph {et~al.}(2017)\citenamefont {Grimm},
		\citenamefont {El{\c c}i}, \citenamefont {Zhou}, \citenamefont {Garoni},\
		and\ \citenamefont {Deng}}]{Grimm2017Geometric}%
	\BibitemOpen
	\bibfield  {author} {\bibinfo {author} {\bibfnamefont {J.}~\bibnamefont
			{Grimm}}, \bibinfo {author} {\bibfnamefont {E.~M.}\ \bibnamefont {El{\c
					c}i}}, \bibinfo {author} {\bibfnamefont {Z.}~\bibnamefont {Zhou}}, \bibinfo
		{author} {\bibfnamefont {T.~M.}\ \bibnamefont {Garoni}},\ and\ \bibinfo
		{author} {\bibfnamefont {Y.}~\bibnamefont {Deng}},\ }\bibfield  {title}
	{\bibinfo {title} {Geometric explanation of anomalous finite-size scaling in
			high dimensions},\ }\href
	{} {\bibfield
		{journal} {\bibinfo  {journal} {Physical Review Letters}\ }\textbf {\bibinfo
			{volume} {118}},\ \bibinfo {pages} {115701} (\bibinfo {year}
		{2017})}\BibitemShut {NoStop}%
	\bibitem [{\citenamefont {Zhou}\ \emph {et~al.}(2018)\citenamefont {Zhou},
		\citenamefont {Grimm}, \citenamefont {Fang}, \citenamefont {Deng},\ and\
		\citenamefont {Garoni}}]{Zhou2018Randomlength}%
	\BibitemOpen
	\bibfield  {author} {\bibinfo {author} {\bibfnamefont {Z.}~\bibnamefont
			{Zhou}}, \bibinfo {author} {\bibfnamefont {J.}~\bibnamefont {Grimm}},
		\bibinfo {author} {\bibfnamefont {S.}~\bibnamefont {Fang}}, \bibinfo {author}
		{\bibfnamefont {Y.}~\bibnamefont {Deng}},\ and\ \bibinfo {author}
		{\bibfnamefont {T.~M.}\ \bibnamefont {Garoni}},\ }\bibfield  {title}
	{\bibinfo {title} {Random-length random walks and finite-size scaling in high
			dimensions},\ }\href
	{} {\bibfield
		{journal} {\bibinfo  {journal} {Physical Review Letters}\ }\textbf {\bibinfo
			{volume} {121}},\ \bibinfo {pages} {185701} (\bibinfo {year}
		{2018})}\BibitemShut {NoStop}%
	\bibitem [{\citenamefont {{J. C. Grimm}}(2018)}]{Grimm2018Finitesize}%
	\BibitemOpen
	\bibfield  {author} {\bibinfo {author} {\bibnamefont {{J. C. Grimm}}},\
	}\emph {\bibinfo {title} {{Finite-size effects in high dimensional physical
				systems}}},\ \href@noop {} {Ph.D. thesis},\ \bibinfo  {school} {{Monash
			University}}, \bibinfo {address} {{Clayton, Victoria, Australia}} (\bibinfo
	{year} {2018})\BibitemShut {NoStop}%
	\bibitem [{\citenamefont {Fang}\ \emph {et~al.}(2020)\citenamefont {Fang},
		\citenamefont {Grimm}, \citenamefont {Zhou},\ and\ \citenamefont
		{Deng}}]{Fang2020Complete}%
	\BibitemOpen
	\bibfield  {author} {\bibinfo {author} {\bibfnamefont {S.}~\bibnamefont
			{Fang}}, \bibinfo {author} {\bibfnamefont {J.}~\bibnamefont {Grimm}},
		\bibinfo {author} {\bibfnamefont {Z.}~\bibnamefont {Zhou}},\ and\ \bibinfo
		{author} {\bibfnamefont {Y.}~\bibnamefont {Deng}},\ }\bibfield  {title}
	{\bibinfo {title} {Complete graph and {Gaussian} fixed-point asymptotics in
			the five-dimensional {Fortuin-Kasteleyn Ising} model with periodic
			boundaries},\ }\href {}
	{\bibfield  {journal} {\bibinfo  {journal} {Physical Review E}\ }\textbf
		{\bibinfo {volume} {102}},\ \bibinfo {pages} {022125} (\bibinfo {year}
		{2020})}\BibitemShut {NoStop}%
	\bibitem [{\citenamefont {Lv}\ \emph {et~al.}(2020)\citenamefont {Lv},
		\citenamefont {Xu}, \citenamefont {Sun}, \citenamefont {Chen},\ and\
		\citenamefont {Deng}}]{Lv2020Two}%
	\BibitemOpen
	\bibfield  {author} {\bibinfo {author} {\bibfnamefont {J.-P.}\ \bibnamefont
			{Lv}}, \bibinfo {author} {\bibfnamefont {W.}~\bibnamefont {Xu}}, \bibinfo
		{author} {\bibfnamefont {Y.}~\bibnamefont {Sun}}, \bibinfo {author}
		{\bibfnamefont {K.}~\bibnamefont {Chen}},\ and\ \bibinfo {author}
		{\bibfnamefont {Y.}~\bibnamefont {Deng}},\ }\bibfield  {title} {\bibinfo
		{title} {Finite-size scaling of {O}(n) systems at the upper critical
			dimensionality},\ }\bibfield  {journal} {\bibinfo  {journal} {National
			Science Review}\ }\href {}
	{10.1093/nsr/nwaa212} (\bibinfo {year} {2020})\BibitemShut {NoStop}%
	\bibitem [{\citenamefont {Camia}\ \emph {et~al.}(2020)\citenamefont {Camia},
		\citenamefont {Jiang},\ and\ \citenamefont {Newman}}]{Camia2020effect}%
	\BibitemOpen
	\bibfield  {author} {\bibinfo {author} {\bibfnamefont {F.}~\bibnamefont
			{Camia}}, \bibinfo {author} {\bibfnamefont {J.}~\bibnamefont {Jiang}},\ and\
		\bibinfo {author} {\bibfnamefont {C.~M.}\ \bibnamefont {Newman}},\ }\href
	{} {\bibinfo {title} {The effect of free
			boundary conditions on the {Ising} model in high dimensions}} (\bibinfo {year}
	{2020}),\ \Eprint {} {arXiv:2011.02814}
	\BibitemShut {NoStop}%
	\bibitem [{\citenamefont {Luczak}\ and\ \citenamefont
		{{\L}uczak}(2006)}]{Luczak2006Phase}%
	\BibitemOpen
	\bibfield  {author} {\bibinfo {author} {\bibfnamefont {M.}~\bibnamefont
			{Luczak}}\ and\ \bibinfo {author} {\bibfnamefont {T.}~\bibnamefont
			{{\L}uczak}},\ }\bibfield  {title} {\bibinfo {title} {The phase transition in
			the cluster-scaled model of a random graph},\ }\href
	{} {\bibfield
		{journal} {\bibinfo  {journal} {Random Structures \& Algorithms}\ }\textbf
		{\bibinfo {volume} {28}},\ \bibinfo {pages} {215} (\bibinfo {year}
		{2006})}\BibitemShut {NoStop}%
	\bibitem [{\citenamefont {Bollob{\'a}s}\ \emph {et~al.}(1996)\citenamefont
		{Bollob{\'a}s}, \citenamefont {Grimmett},\ and\ \citenamefont
		{Janson}}]{Bollobas1996randomcluster}%
	\BibitemOpen
	\bibfield  {author} {\bibinfo {author} {\bibfnamefont {B.}~\bibnamefont
			{Bollob{\'a}s}}, \bibinfo {author} {\bibfnamefont {G.}~\bibnamefont
			{Grimmett}},\ and\ \bibinfo {author} {\bibfnamefont {S.}~\bibnamefont
			{Janson}},\ }\bibfield  {title} {\bibinfo {title} {The random-cluster model
			on the complete graph},\ }\href
	{} {\bibfield  {journal}
		{\bibinfo  {journal} {Probability Theory and Related Fields}\ }\textbf
		{\bibinfo {volume} {104}},\ \bibinfo {pages} {283} (\bibinfo {year}
		{1996})}\BibitemShut {NoStop}%
	\bibitem [{\citenamefont {Fang}\ \emph {et~al.}(2021)\citenamefont {Fang},
		\citenamefont {Zhou},\ and\ \citenamefont {Deng}}]{Fang2021Percolation}%
	\BibitemOpen
	\bibfield  {author} {\bibinfo {author} {\bibfnamefont {S.}~\bibnamefont
			{Fang}}, \bibinfo {author} {\bibfnamefont {Z.}~\bibnamefont {Zhou}},\ and\
		\bibinfo {author} {\bibfnamefont {Y.}~\bibnamefont {Deng}},\ }\bibfield
	{title} {\bibinfo {title} {Percolation effects in the {Fortuin-Kasteleyn
				Ising} model on the complete graph},\ }\href
	{} {\bibfield  {journal}
		{\bibinfo  {journal} {Physical Review E}\ }\textbf {\bibinfo {volume}
			{103}},\ \bibinfo {pages} {012102} (\bibinfo {year} {2021})}\BibitemShut
	{NoStop}%
	\bibitem [{\citenamefont {{Ben-Naim}}\ and\ \citenamefont
		{Krapivsky}(2005)}]{Ben-Naim2005Kinetica}%
	\BibitemOpen
	\bibfield  {author} {\bibinfo {author} {\bibfnamefont {E.}~\bibnamefont
			{{Ben-Naim}}}\ and\ \bibinfo {author} {\bibfnamefont {P.~L.}\ \bibnamefont
			{Krapivsky}},\ }\bibfield  {title} {\bibinfo {title} {Kinetic theory of
			random graphs: From paths to cycles},\ }\href
	{} {\bibfield  {journal}
		{\bibinfo  {journal} {Physical Review E}\ }\textbf {\bibinfo {volume} {71}},\
		\bibinfo {pages} {026129} (\bibinfo {year} {2005})}\BibitemShut {NoStop}%
	\bibitem [{\citenamefont {Aharony}\ \emph {et~al.}(1984)\citenamefont
		{Aharony}, \citenamefont {Gefen},\ and\ \citenamefont
		{Kapitulnik}}]{Aharony1984Scaling}%
	\BibitemOpen
	\bibfield  {author} {\bibinfo {author} {\bibfnamefont {A.}~\bibnamefont
			{Aharony}}, \bibinfo {author} {\bibfnamefont {Y.}~\bibnamefont {Gefen}},\
		and\ \bibinfo {author} {\bibfnamefont {A.}~\bibnamefont {Kapitulnik}},\
	}\bibfield  {title} {\bibinfo {title} {Scaling at the percolation threshold
			above six dimensions},\ }\href
	{} {\bibfield
		{journal} {\bibinfo  {journal} {Journal of Physics A: Mathematical and
				General}\ }\textbf {\bibinfo {volume} {17}},\ \bibinfo {pages} {L197}
		(\bibinfo {year} {1984})}\BibitemShut {NoStop}%
	\bibitem [{\citenamefont {Lu}\ \emph {et~al.}(2023)\citenamefont {Lu},
		\citenamefont {Fang}, \citenamefont {Zhou},\ and\ \citenamefont
		{Deng}}]{Lu2023Finite}%
	\BibitemOpen
	\bibfield  {author} {\bibinfo {author} {\bibfnamefont {M.}~\bibnamefont
			{Lu}}, \bibinfo {author} {\bibfnamefont {S.}~\bibnamefont {Fang}}, \bibinfo
		{author} {\bibfnamefont {Z.}~\bibnamefont {Zhou}},\ and\ \bibinfo {author}
		{\bibfnamefont {Y.}~\bibnamefont {Deng}},\ }\href@noop {} {\bibfield
		{journal} {\bibinfo  {journal} {in preparation}\ } (\bibinfo {year}
		{2023})}\BibitemShut {NoStop}%
	\bibitem [{\citenamefont {Huang}\ \emph {et~al.}(2018)\citenamefont {Huang},
		\citenamefont {Hou}, \citenamefont {Wang}, \citenamefont {Ziff},\ and\
		\citenamefont {Deng}}]{Huang2018Critical}%
	\BibitemOpen
	\bibfield  {author} {\bibinfo {author} {\bibfnamefont {W.}~\bibnamefont
			{Huang}}, \bibinfo {author} {\bibfnamefont {P.}~\bibnamefont {Hou}}, \bibinfo
		{author} {\bibfnamefont {J.}~\bibnamefont {Wang}}, \bibinfo {author}
		{\bibfnamefont {R.~M.}\ \bibnamefont {Ziff}},\ and\ \bibinfo {author}
		{\bibfnamefont {Y.}~\bibnamefont {Deng}},\ }\bibfield  {title} {\bibinfo
		{title} {Critical percolation clusters in seven dimensions and on a complete
			graph},\ }\href{}
	{\bibfield  {journal} {\bibinfo  {journal} {Physical Review E}\ }\textbf
		{\bibinfo {volume} {97}},\ \bibinfo {pages} {022107} (\bibinfo {year}
		{2018})}\BibitemShut {NoStop}%
	\bibitem [{\citenamefont {Aizenman}(1997)}]{Aizenman1997number}%
	\BibitemOpen
	\bibfield  {author} {\bibinfo {author} {\bibfnamefont {M.}~\bibnamefont
			{Aizenman}},\ }\bibfield  {title} {\bibinfo {title} {On the number of
			incipient spanning clusters},\ }\href
	{} {\bibfield
		{journal} {\bibinfo  {journal} {Nuclear Physics B}\ }\textbf {\bibinfo
			{volume} {485}},\ \bibinfo {pages} {551} (\bibinfo {year}
		{1997})}\BibitemShut {NoStop}%
	\bibitem [{\citenamefont {Fortunato}\ \emph {et~al.}(2004)\citenamefont
		{Fortunato}, \citenamefont {Aharony}, \citenamefont {Coniglio},\ and\
		\citenamefont {Stauffer}}]{Fortunato2004Number}%
	\BibitemOpen
	\bibfield  {author} {\bibinfo {author} {\bibfnamefont {S.}~\bibnamefont
			{Fortunato}}, \bibinfo {author} {\bibfnamefont {A.}~\bibnamefont {Aharony}},
		\bibinfo {author} {\bibfnamefont {A.}~\bibnamefont {Coniglio}},\ and\
		\bibinfo {author} {\bibfnamefont {D.}~\bibnamefont {Stauffer}},\ }\bibfield
	{title} {\bibinfo {title} {Number of spanning clusters at the
			high-dimensional percolation thresholds},\ }\href
	{} {\bibfield  {journal}
		{\bibinfo  {journal} {Physical Review E}\ }\textbf {\bibinfo {volume} {70}},\
		\bibinfo {pages} {056116} (\bibinfo {year} {2004})}\BibitemShut {NoStop}%
	\bibitem [{\citenamefont {Bl{\"o}te}\ and\ \citenamefont
		{Luijten}(1997)}]{Blote1997Universality}%
	\BibitemOpen
	\bibfield  {author} {\bibinfo {author} {\bibfnamefont {H.~W.~J.}\
			\bibnamefont {Bl{\"o}te}}\ and\ \bibinfo {author} {\bibfnamefont
			{E.}~\bibnamefont {Luijten}},\ }\bibfield  {title} {\bibinfo {title}
		{Universality and the five-dimensional {Ising} model},\ }\href
	{} {\bibfield
		{journal} {\bibinfo  {journal} {Europhysics Letters (EPL)}\ }\textbf
		{\bibinfo {volume} {38}},\ \bibinfo {pages} {565} (\bibinfo {year}
		{1997})}\BibitemShut {NoStop}%
	\bibitem [{\citenamefont {Lundow}\ and\ \citenamefont
		{Markstr{\"o}m}(2015{\natexlab{a}})}]{Lundow2015Discontinuity}%
	\BibitemOpen
	\bibfield  {author} {\bibinfo {author} {\bibfnamefont {P.}~\bibnamefont
			{Lundow}}\ and\ \bibinfo {author} {\bibfnamefont {K.}~\bibnamefont
			{Markstr{\"o}m}},\ }\bibfield  {title} {\bibinfo {title} {The discontinuity
			of the specific heat for the {5D Ising} model},\ }\href
	{} {\bibfield
		{journal} {\bibinfo  {journal} {Nuclear Physics B}\ }\textbf {\bibinfo
			{volume} {895}},\ \bibinfo {pages} {305} (\bibinfo {year}
		{2015}{\natexlab{a}})}\BibitemShut {NoStop}%
	\bibitem [{\citenamefont {Lundow}\ and\ \citenamefont
		{Markstr{\"o}m}(2015{\natexlab{b}})}]{Lundow2015Complete}%
	\BibitemOpen
	\bibfield  {author} {\bibinfo {author} {\bibfnamefont {P.~H.}\ \bibnamefont
			{Lundow}}\ and\ \bibinfo {author} {\bibfnamefont {K.}~\bibnamefont
			{Markstr{\"o}m}},\ }\bibfield  {title} {\bibinfo {title} {Complete graph
			asymptotics for the {Ising} and random-cluster models on five-dimensional
			grids with a cyclic boundary},\ }\href
	{} {\bibfield  {journal}
		{\bibinfo  {journal} {Physical Review E}\ }\textbf {\bibinfo {volume} {91}},\
		\bibinfo {pages} {022112} (\bibinfo {year} {2015}{\natexlab{b}})}\BibitemShut
	{NoStop}%
	\bibitem [{\citenamefont {Luijten}(1997)}]{Luijten1997Interaction}%
	\BibitemOpen
	\bibfield  {author} {\bibinfo {author} {\bibfnamefont {E.}~\bibnamefont
			{Luijten}},\ }\href@noop {} {\emph {\bibinfo {title} {Interaction Range,
				Universality and the Upper Critical Dimension}}}\ (\bibinfo  {publisher}
	{Delft Univ. Press},\ \bibinfo {address} {Delft},\ \bibinfo {year}
	{1997})\BibitemShut {NoStop}%
	\bibitem [{\citenamefont {Li}\ \emph {et~al.}(2023)\citenamefont {Li},
		\citenamefont {Zhou}, \citenamefont {Fang},\ and\ \citenamefont
		{Deng}}]{Li2023Geometric}%
	\BibitemOpen
	\bibfield  {author} {\bibinfo {author} {\bibfnamefont {Z.}~\bibnamefont
			{Li}}, \bibinfo {author} {\bibfnamefont {Z.}~\bibnamefont {Zhou}}, \bibinfo
		{author} {\bibfnamefont {S.}~\bibnamefont {Fang}},\ and\ \bibinfo {author}
		{\bibfnamefont {Y.}~\bibnamefont {Deng}},\ }\href
	{} {\bibinfo {title} {Geometric properties of
			the complete-graph {Ising} model in the loop representation}} (\bibinfo {year}
	{2023}),\ \Eprint {} {arXiv:2302.14381} \BibitemShut {NoStop}%
\end{thebibliography}
%

\end{document}